\newif\ifconfver
\confvertrue       
\ifconfver
\documentclass[10pt,twocolumn,twoside]{IEEEtran}
\else
\documentclass[11pt,draftcls,onecolumn]{IEEEtran}
\fi

\usepackage{makecell}
\usepackage{bbding}
\usepackage{color,graphicx}
\usepackage{float}
\usepackage{multirow,amsmath,epsfig,amsfonts,amssymb,psfig,graphics,psfrag,theorem,calc,url,bm,cite}
\usepackage{stfloats,hyperref}
\usepackage{algorithm,algorithmic}
\usepackage{diagbox} 
\usepackage{hhline}
\usepackage{subfigure}
\usepackage{physics}
\usepackage{booktabs}
\usepackage{tikz}
\usepackage{multirow}
\usepackage{arydshln}
\usepackage{mathtools}

\makeatletter
\def\multilimits@{\bgroup
	\Let@
	\restore@math@cr
	\default@tag
	\baselineskip\fontdimen10 \scriptfont\tw@
	\advance\baselineskip\fontdimen12 \scriptfont\tw@
	\lineskip\thr@@\fontdimen8 \scriptfont\thr@@
	\lineskiplimit\lineskip
	\vbox\bgroup\ialign\bgroup\hfil$\m@th\scriptstyle{##}$\hfil\crcr}
\def\Sb{_\multilimits@}
\def\endSb{\crcr\egroup\egroup\egroup}
\makeatother

	{\end{list}}

\newlength{\twidth}
\ifconfver
\else
\fi

\definecolor{orange}{RGB}{255,107,0}

\theorembodyfont{\rmfamily}

{\begin{list}{}{
			\settowidth{\labelwidth}{\mbox{\textnormal{#1}}}%
			\setlength{\leftmargin}{\labelwidth+\labelsep}}}%
	{\end{list}}

\newcommand\bD{\ensuremath{{\bm D}}}

\newcommand\bI{\ensuremath{{\bm I}}}

\newcommand\bQ{\ensuremath{{\bm Q}}}

\newcommand\bU{\ensuremath{{\bm U}}}
\newcommand\bV{\ensuremath{{\bm V}}}

\newcommand\bX{\ensuremath{{\bm X}}}
\newcommand\bY{\ensuremath{{\bm Y}}}
\newcommand\bZ{\ensuremath{{\bm Z}}}

\newcommand\bv{\ensuremath{{\bm v}}}
\newcommand\bw{\ensuremath{{\bm w}}}

\usepackage{listings}
\usepackage{xcolor}

\definecolor{orange}{RGB}{255,107,0}

\ifconfver

\author{Chia-Hsiang Lin,~\IEEEmembership{Senior Member,~IEEE}, and Zi-Chao Leng,~\IEEEmembership{Graduate Student Member,~IEEE}}

\title{ExplainS2A: Explainable Spectral-Spatial Duality Model for Fast Transforming Sentinel-2 Image to AVIRIS-Level Hyperspectral Image

\thanks{This study was supported by the Emerging Young Scholar Program (namely, the 2030 Cross-Generation Young Scholars Program) of National Science and Technology Council (NSTC), Taiwan, under Grant NSTC 114-2628-E-006-002.
We thank the National Center for Theoretical Sciences (NCTS) and the National Center for High-performance Computing (NCHC) for providing the computing resources.
\textit{(Corresponding Author: Chia-Hsiang Lin)}}
\thanks{C.-H. Lin is with the Department of Electrical Engineering, National Cheng Kung University, Tainan 70101, Taiwan 
(e-mail: chiahsiang.steven.lin@gmail.com).}
\thanks{Z.-C. Leng is with the Institute of Computer and Communication Engineering, Department of Electrical Engineering, National Cheng Kung University, Tainan 70101, Taiwan 
(e-mail:  q38115558@gs.ncku.edu.tw).}
}
\else

\fi

\begin{document}
	
	\bibliographystyle{IEEEtran}
	\maketitle
	\ifconfver \else \vspace{-0.5cm}\fi

\begin{abstract}
Mainstream optical satellites (e.g., ESA's Sentinel-2) often acquire multispectral multi-resolution images, which have limited material identifiability compared to the hyperspectral images (HSI).
Thus, spectrally super-resolving the multispectral image (MSI) into their hyperspectral counterparts greatly facilitates remote material identification and the downstream tasks.
However, spectrally super-resolving the MSI into an HSI is often constrained by the multi-resolution nature of the sensor (e.g., Sentinel-2).
%
%
Specifically, due to the presence of some low-resolution (LR) bands in the MSI, the initial spectral super-resolution results often appear to be spatially blurry, resulting in an LR HSI.
To overcome this bottleneck, we then leverage some high-resolution (HR) band inherent in the acquired MSI (e.g., panchromatic band) to spatially guide the reconstruction procedure, thereby yielding the desired HR HSI.
This fusion procedure elegantly coincides with a widely known spatial super-resolution problem in satellite remote sensing.
Hence, we have reformulated the tough spectral super-resolution problem into a more widely investigated spatial super-resolution problem, referred to as the spectral-spatial duality theory.
Accordingly, we propose ExplainS2A, consisting of a deep unfolding network and an explainable fusion network, that unifies spectral recovery and spatial fusion into a single explainable framework.
Unlike conventional black-box models, ExplainS2A offers interpretability and operates as a linear-time algorithm.
%
Remarkably, it can process a million-scale Sentinel-2 image in less than one second, yielding high-fidelity HSI over the same scene, and upgrades the blind source separation results.
Although demonstrated on the Sentinel-2 and AVIRIS sensors, ExplainS2A also serves as a general framework applicable to various sensor pairs with different resolution configurations, and has experimentally demonstrated cross-region and cross-season generalization ability. 
Source codes: \url{https://github.com/IHCLab/ExplainS2A}.
\end{abstract}

\begin{IEEEkeywords}
Sentinel-2 satellite,
AVIRIS image,
hyperspectral image,
multi-resolution image,
spectral super-resolution,
image fusion,
explainable AI (XAI).
\end{IEEEkeywords}
	
	\ifconfver \else \vspace{-0.0cm}\fi
	
	\ifconfver \else \vspace{-0.5cm}\fi
	
	\ifconfver \else  \fi

\section{Introduction}\label{sec: introduction}

Although mainstream optical satellites provide critical Earth observation data, they often acquire multispectral imagery with varying spatial resolutions, which lacks the rich spectral information of hyperspectral images (HSIs) for precise and provable material identifiability \cite{EMI}.
%
For example, the Sentinel-2 satellite launched by European Space Agency (ESA) has only 4 high-resolution (HR) bands with 10-m ground sample distance (GSD), 6 medium-resolution bands with 20-m GSD, and 3 low-resolution (LR) bands with 60-m GSD \cite{drusch2012sentinel}.
As one of the LR bands (i.e., band 10) contains only the cirrus information for atmospheric correction, there are actually only 12 multispectral multi-resolution bands in Sentinel-2.
To computationally upgrade the material identifiability, spectrally super-resolving the multispectral images (MSIs) into their hyperspectral counterparts becomes a critical and high-impact signal processing technology.
This greatly facilitates remote material identification and classification, as well as the downstream tasks and applications.
This work ambitiously aims to transform Sentinel-2 data directly into the AVIRIS-level HSI, where AVIRIS data encompasses up to 172 visible and near-infrared (NIR) bands, after removing those low-quality absorption/corruption bands  \cite{AVIRISdata}.
As Sentinel-2 offers wide spatiotemporal coverages, having become indispensable in various remote sensing applications and sustainable development goals (SDGs) \cite{FTDN2025,phiri2020sentinel,segarra2020remote,CVPR2026}, the impact of our investigated spectral super-resolution technique is foreseeable.
It also enables retrospective construction of high-fidelity AVIRIS-like hyperspectral data over regions previously covered by Sentinel-2, thus greatly expanding the utility of existing Earth observation archives.

However, owing to fundamental hardware limitations of imaging sensors, as well as practical constraints such as atmospheric absorption and the necessity of maintaining an adequate signal-to-noise ratio (SNR), the spatial resolution of Sentinel-2 imagery exhibits variability across different spectral bands, as aforementioned.
%
%
While several spectral super-resolution methods have been proposed in recent literature, most existing approaches do not consider the aforementioned multi-resolution nature of the target problem.
More seriously, these methods primarily target benchmark-level hyperspectral reconstruction tasks (e.g., Harvard/CAVE), which are restricted to only 31 spectral bands in the visible range \cite{LTRN,CAVEdata,he2023spectral}.
Such constraints render these methods inadequate for operational remote sensing tasks, which often rely on near-infrared (NIR) bands for improved atmospheric penetration, material discrimination, and vegetation monitoring \cite{WOL2017significant,FTDN2025}.
To overcome these shortcomings, we aim to transform Sentinel-2 multispectral imagery into AVIRIS-grade hyperspectral data composed of up to 172 contiguous and narrowly spaced bands spanning the visible to shortwave infrared (SWIR) region, from 0.4 to 2.5 $\mu$m.
This amounts to a highly challenging inverse problem.

Despite the target spectral super-resolution task being a severely ill-posed inverse problem, emerging evidence suggests its technical viability. 
It has been shown to be feasible not only in data-driven software approaches but also within hyperspectral hardware systems \cite{NCCODE}.
For example, to meet the strict payload constraints of miniaturized satellite platforms \cite{ferre2022feasibility,AAHCSD}, compact hyperspectral meta-imaging systems using flat, nanoscale metamaterials \cite{wang2024advances} have been proposed recently.
These systems first acquire a low-spectral-resolution meta-image, which is then computationally upsampled to generate a hyperspectral cube \cite{NCCODE}, demonstrating the real-world potential of our target problem under resource-constrained environments.
On the algorithmic front, recent advances such as the quantum deep network (QUEEN) \cite{HyperQUEEN,QUEENG} have been developed to tackle underdetermined spectral unmixing problems.
Specifically, QUEEN functions analogously to a virtual quantum light-splitting prism, leveraging quantum deep image prior (QDIP) \cite{PRIME,TIP2026quantumMU} to spectrally super-resolve the input spectra by splitting coarse multispectral inputs into fine-grained hyperspectral bands.
Nevertheless, despite these innovations, current techniques are constrained to modest spectral upsampling factors \cite{NCCODE,LTRN,MST++}, which are insufficient for the demands of the target task, where the goal is to scale from 12-band data to over 170 hyperspectral bands.
This highly challenging task has previously been achieved only once through the algorithm called COS2A (i.e., ``conversion from Sentinel-2 to AVIRIS'') \cite{COS2A}, albeit not in a computationally cheap manner.

The COS2A algorithm is developed based on the so-called convex/deep (CODE) learning theory \cite{CODE}, which was originally customized for solving challenging hyperspectral inverse imaging problems.
The CODE learning has led to numerous state-of-the-art remote sensing software, including those for blue carbon ecosystem monitoring \cite{CODEMM}, land cover change detection \cite{CODEHCD}, and satellite image fusion \cite{CODEIF}.
It has also been applied to the 31-band spectral super-resolution task \cite{lin2022fastRGB}, exhibiting strong performance across various benchmarks.
The CODE theory blends the advantages from the alternating direction method of multipliers (ADMM) in convex optimization \cite{CVXbookCLL2016} and adaptive moment estimation (ADAM) in deep learning \cite{kingma2014adam} using $\bQ$-quadratic-norm regularization scheme, thereby avoiding math-heavy derivation in convex optimization and the big data requirement in deep learning.
Accordingly, the COS2A algorithm \cite{COS2A} considers the deep learning solution just as a regularization role, allowing the deep phase just to return a rough solution (learned from small data), thereby greatly mitigating the burden of the deep learning phase.
The rough deep solution is then plugged into a simple convex optimization problem with a $\bQ$-quadratic-norm regularizer, which is solved using ADMM to complete the spectral super-resolution mission.
Through the CODE learning theory, the COS2A algorithm transforms the highly ill-posed spectral super-resolution problem into a simple convex problem, thereby elegantly solving it using simple matrix factorizations \cite{COS2A}.
To the best of our knowledge, COS2A is the only algorithm capable of effectively transforming Sentinel-2 data into AVIRIS-grade hyperspectral imagery.

Another seminal spectral super-resolution algorithm, called multi-stage spectral-wise transformer (MST++) \cite{MST++}, which represents a significant advancement in transformer-based hyperspectral reconstruction in the visible wavelength region.
Although MST is also designed for the 31-band visible spectrum, it demonstrates exceptional performance upon its release and won the first place in the renowned New Trends in Image Restoration and Enhancement (NTIRE) Challenge.
MST introduces a spectral-wise attention mechanism by modeling each spectral band as an individual token, thereby capturing long-range spectral dependencies. 
Its multi-stage architecture also enables the network to progressively refine spectral representations at increasing levels of detail, contributing to its outstanding performance.
As a side remark, state-space-model-based architectures, such as Mamba \cite{CR-Famba, SQUARE-Mamba}, have emerged as an efficient alternative to transformer-based designs. 
Mamba enables long-range dependency modeling with linear computational complexity, while avoiding the quadratic overhead inherent to the self-attention mechanism.
In parallel with supervised methods, the unmixing guided unsupervised network (UnGUN) \cite{UnGUN} performs spectral super-resolution without pairwise imagery by unmixing an RGB image and a guidance HSI to extract spectral-spatial priors.
Beyond 31-band HSI reconstruction, some recent works have sought to extend super-resolution capabilities toward higher spectral dimensionality, targeting up to 102 HSI bands. 
Notable examples include the spectral-cascade diffusion model (SCDM) \cite{SCDM} and the unfolding spatiospectral super-resolution network (US3RN) \cite{deepUnfolding}.
SCDM employs a probabilistic diffusion framework to model the spectral distribution and applies multi-stage refinement to iteratively enhance band-level quality. 
In contrast, US3RN reformulates the target task into a structured optimization problem and then solves it via deep unfolding.
We should also mention some recent architectures, including the multistage spatial-spectral fusion network (MSFN) \cite{MSFN}, the reparameterized coordinate-preserving proximity spectral interaction network (RepCPSI) \cite{RepCPSI}, and the spectral-spatial residual attention U-Net (SSU-Net) \cite{SSU-Net}.
%
MSFN utilizes a U-Net-inspired CNN backbone to jointly model multi-scale spatial context and spectral self-similarity, facilitating effective spatial-spectral feature learning.
RepCPSI, on the other hand, leverages a multi-branch convolutional strategy to improve spectral detail modeling during training. 
Importantly, it introduces a coordinate-preserving attention module, which retains spatial positional integrity while capturing localized spectral dependencies, resulting in a lightweight network design.
SSU-Net employs a weight-adaptive allocation module to mitigate inconsistent spectral reconstruction induced by spatially varying land-cover distributions. 
Furthermore, it designs a spectral-spatial residual attention module to capture both long-range spectral dependencies and local spatial correlations. 
Although SSU-Net handles the reconstruction with up to 289 bands, it does not account for a multi-resolution scenario.
%

However, given the multi-resolution nature of Sentinel-2 imagery, directly performing spectral super-resolution inevitably yields spatially blurry output, resulting in an LR HSI.
To tackle this dilemma, we then leverage the HR bands inherent in the acquired MSI to facilitate spatial texture reconstruction of the LR HSI, thereby yielding the desired AVIRIS-level HR HSI.
%
The above fusion strategy elegantly coincides with a widely known spatial super-resolution problem in satellite remote sensing, often formulated using coupled matrix factorizations \cite{COCNMF}.
In other words, we have reformulated the highly ill-posed spectral super-resolution problem into a more widely investigated spatial super-resolution problem, and we refer to it as the spectral-spatial duality theory.
Accordingly, we propose a deep unfolding network to form the above LR HSI, and design an explainable network to conduct the fusion task, thereby leading to an explainable and linear-time algorithm, termed as ExplainS2A, which can effectively and efficiently transform Sentinel-2 data into high-standard AVIRIS hyperspectral data.
Remarkably, the proposed ExplainS2A algorithm is able to process a million-scale Sentinel-2 image in less than one second, yielding a high-fidelity HSI for the subsequent remote sensing tasks such as blind source separation.
While this article focuses on the Sentinel-2-to-AVIRIS task, the ExplainS2A theory is generally applicable to various MSI/HSI sensor pairs with different resolution configurations.
The novelty and main contributions are itemized below.
\begin{enumerate}
\item 
Typical spectral super-resolution models are often restricted to recovering only 31 spectral bands in the visible range from single-resolution inputs (e.g., RGB images), while the proposed ExplainS2A algorithm addresses the more challenging multi-resolution scenario.
Furthermore,  ExplainS2A transforms Sentinel-2 MSIs into AVIRIS-grade HSIs composed of up to 172 bands spanning the visible to NIR ranges.
This represents a highly challenging ill-posed inverse problem, as it requires handling varying spatial resolutions while achieving up to about 15 times spectral super-resolution with state-of-the-art performance.

\item 
Unlike conventional spectral super-resolution models that operate as uninterpretable black boxes, the proposed ExplainS2A consists of explainable network architectures, thereby effectively solving the challenging ill-posed problem.
Specifically, ExplainS2A first employs an ADMM-driven deep unfolding network to complete the spectral super-resolution mission.
Due to the multi-resolution nature of the input, the result inevitably remains an LR HSI (as introduced above), motivating us to adopt the spectral-spatial duality discovered in very recent remote sensing literature \cite{COS2A}.
However, the existing method for implementing the duality is computationally heavy \cite{COS2A}.
Therefore, our proposed explainable network not only learns explicit spectral-spatial attentions but also implements the spectral-spatial duality to reconstruct the target HSI in a real-time manner.

\item  
Compared to the only existing Sentinel-to-AVIRIS reconstruction method, i.e., the CODE-based COS2A algorithm \cite{COS2A}, which suffers from heavy computational burdens, ExplainS2A achieves a breakthrough in both efficiency and interpretability, besides the rather superior spectral super-resolution performance.
In detail, it features linear computational complexity and can process a million-scale Sentinel-2 image in less than one second (cf. Figure \ref{fig:linear}).
In addition, ExplainS2A maintains high spectral reconstruction fidelity, which contributes most to the outstanding material identifiability of HSI.
We also demonstrate that the spectral fidelity does well to upgrade the performance of downstream tasks, such as blind source separation (cf. Section \ref{sec:CaseStudy}).
Furthermore, the theory behind ExplainS2A is more practically applicable (as it considers the more frequently encountered multi-resolution nature of satellite MSI), and serves as a more general framework applicable to various sensor pairs with different resolution configurations (having experimentally demonstrable cross-region and cross-season model generalization ability).
\end{enumerate}

The remaining parts of this paper are organized as follows.
In Section \ref{sec:thm}, we present the design of the proposed ExplainS2A algorithm, including the problem formulation and network architectures.
In Section \ref{sec: experiment}, we experimentally demonstrate the strong performance and computational efficiency of ExplainS2A.
In Section \ref{sec:conclusion}, we conclude this work and propose some future research lines.
%

\section{The Proposed ExplainS2A Algorithm}\label{sec:thm}

\subsection{Problem Description}\label{sec:probdesc}

Denote the multi-resolution Sentinel-2 image as $ \bY_S\in\mathbb{R}^{M_m\times L}$, where $M_m:=12$ is the number of Sentinel-2 spectral bands, and $L$ is the number of pixels in the HR 10-m band.
The multi-resolution Sentinel-2 image can be compactly represented as the matrix $\bY_S$, because those pixels in the medium-resolution 20-m bands are copied 4 times and those in the LR 60-m bands are copied 36 times \cite{SSSS}, following the format as downloaded from ESA's data portal.
Noted that although other upsampling methods (e.g., bicubic interpolation \cite{Han2013}) are also applicable to represent $\bY_S$, the choice of the upsampling method exerts a marginal impact on the final performance. 
This is because the recovery of spatial structures primarily relies on the HR bands of $\bY_S$ (cf. Section \ref{sec:ID}).
The proposed ExplainS2A algorithm is user-friendly as it does not require the users to spatially super-resolve the medium-resolution and LR bands ahead of time.
After ignoring those absorption/corruption bands \cite{CODE}, the ExplainS2A algorithm aims to computationally transform Sentinel-2 multispectral imagery into its counterpart AVIRIS-grade hyperspectral image $\bY_H\in\mathbb{R}^{M \times L}$ composed of up to $M:=172$ contiguous and narrowly spaced bands (each of 10-m resolution), spanning the visible to SWIR region (from 0.4 to 2.5 $\mu$m) for facilitating various remote sensing missions.

This inverse imaging problem exhibits a high degree of ill-posedness and is considered much more challenging than conventional CAVE-level spectral super-resolution tasks, which predominantly aim to reconstruct standard 31-band CAVE HSI over only the visible region \cite{MST++}.
More seriously, the models they adopted are not explainable.
To the best of our knowledge, this level of performance has been achieved only once in the literature by the CODE-based COS2A algorithm \cite{COS2A}, though the computational time is still unsatisfactory.
Also, as Sentinel-2 image is a multi-resolution one, its spectral super-resolution results often appear to be spatially blurred, limiting its practical applications.
In addition, acquiring well-aligned AVIRIS and Sentinel-2 image pairs captured over identical geographical regions and timeframes is notably challenging, due to factors such as disparities in spatial resolution (which are not integer multiples) and inconsistencies in satellite/flight flying trajectories \cite{S2data,AVIRISdata}.
These constraints result in a learning scenario characterized by limited data availability.
Thus, we aim to develop an effective and explainable algorithm (i.e., ExplainS2A) to solve the above challenges, thereby computationally transforming Sentinel-2 data into AVIRIS-level HSI in a near-real-time manner.

\begin{figure}[t]
\centering
\includegraphics[width=1\linewidth]{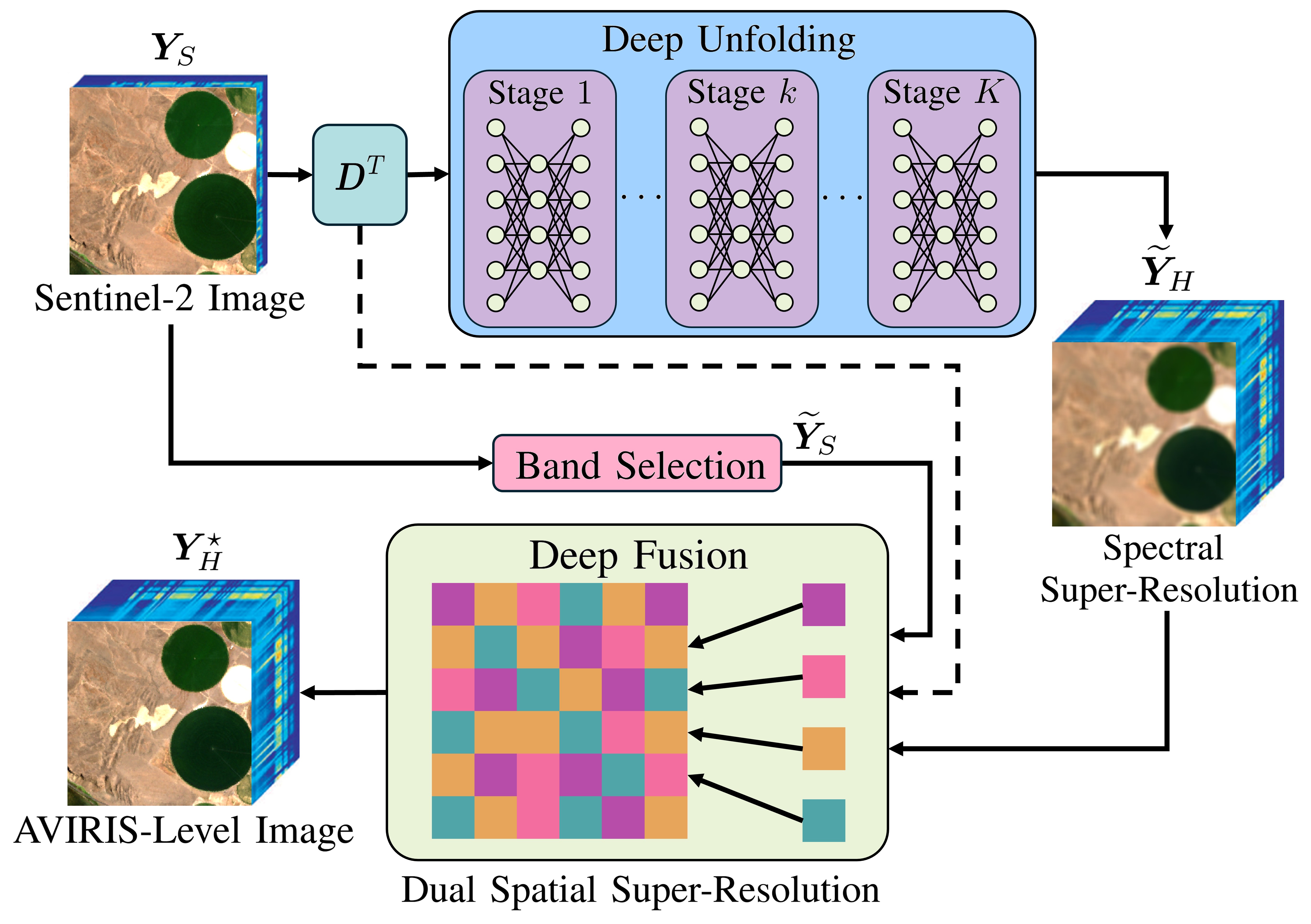}
\vspace{-0.5cm}
\caption{The overview of the proposed ExplainS2A algorithm.
ExplainS2A accepts the 12-band multi-resolution Sentinel-2 image $\bY_S$ as the only input, and then converts it into high-fidelity NASA AVIRIS hyperspectral data $\bY_H^\star$.
The deep unfolding network spectrally super-resolves $\bY_S$ into an intermediate HSI $\widetilde{\bY}_H$, which appeared to be somewhat blurry owing to the multi-resolution nature of the input.
The LR HSI $\widetilde{\bY}_H$ is hence fused with an HR MSI $\widetilde{\bY}_S$ (obtained through HR band selection) according to the spectral-spatial duality theory \cite[Theorem 1]{COS2A}, thereby yielding the desired HR HSI $\bY_H^\star$.
The SRT matrix $\bD$ is set as a learnable parameter initialized with a Xavier normal distribution \cite{glorot2010understanding} during the deep unfolding stage, since no openly available response function is available between Sentinel-2 and AVIRIS sensors, and it propagates to the deep fusion stage through the end-to-end training scheme.
%
%
The unfolding architecture will be detailed in Figure \ref{fig:whole_model}(a), while the fusion network will be designed in Figure \ref{fig:whole_model}(b).
}\label{fig:overview}
\end{figure}

\subsection{Algorithm Design}\label{sec:algodesign}

\begin{figure*}[t!]
\centering
\includegraphics[width=\linewidth]{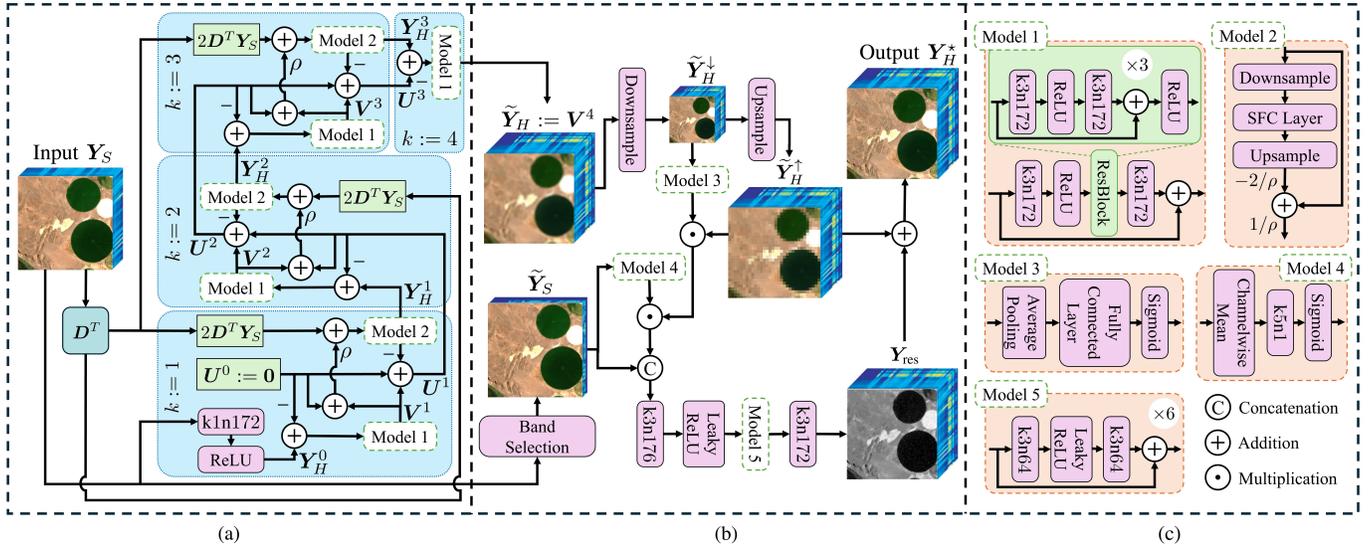}
\vspace{-0.65cm}
\caption{The proposed ExplainS2A spectral super-resolution algorithm for computationally transforming Sentinel-2 image $\bY_S$ to AVIRIS-level hyperspectral data $\bY_H^\star$. 
(a) The ADMM-driven unfolding network requires the initialization of $\bU^0$ and $\bY_H^0$ (before computing $\bV_1$), which are simply initialized as the zero matrix and the spectrally upsampled version of $\bY_S$, respectively.
Besides the initialization part, the first three stages are architecturally identical sharing the same parameter $\bD$, while the last stage ($k:=4$) is a simplified one as it just needs to return $\widetilde{\bY}_H:=\bV^4$ for solving \eqref{prob:DIP-I}.
The network component ``k$(s)$n$(c)$'' is the $c$-channel convolution layer with $s\times s$ kernels \cite{CODE}, while $\rho>0$ is a learnable penalty parameter.
Model 1 and Model 2 are to implement \eqref{eq:V-update} and \eqref{eq:Y-update}, respectively.
(b) The deep fusion network further enhances the spatial quality of $\widetilde{\bY}_H$ by fusing it with the HR data $\widetilde{\bY}_S$, according to the spectral-spatial duality theory, for obtaining the target HSI $\bY_H^\star$.
Model 3 and Model 4 are to implement \eqref{eq:wspec} and \eqref{eq:wspat}, respectively, for emphasizing the key spectral and spatial components, thereby facilitating the learning efficiency of Model 5 that aims to recover missing spatiospectral detail $\bY_\textrm{res}$.
(c) Some necessary network models are collectively defined for better readability and reproducibility.
}
\label{fig:whole_model}
\end{figure*}

The ill-posed spectral super-resolution problem can be naturally formulated as
\begin{equation}\label{prob:DIP}
\widetilde{\bY}_H
:=
\arg\min_{\bY_H}\|\bY_S-\bD\bY_H\|_F^2+\textrm{REG}(\bY_H),
\end{equation}
where $\bD\in\mathbb{R}^{M_m\times M}$ is the spectral response transform (SRT) matrix, $\|\cdot\|_F$ is the Frobenius norm, and $\textrm{REG}(\cdot)$ is to regularize the target ill-posed problem.
Specifically, each row of $\bD$ corresponds to the spectral response of a specific MSI band \cite{COCNMF, 7025846, QRCODE}.
As the exact spectral response specifications between Sentinel-2 and AVIRIS sensors have not been officially released, $\bD$ must be estimated (cf. Figure \ref{fig:overview}).
To facilitate a user-friendly mechanism, we set $\bD$ as a learnable parameter, initialized with a Xavier normal distribution \cite{glorot2010understanding} and shared across stages in our explainable unfolding network, as will be designed in Figure \ref{fig:whole_model}.
The regularization function involved in \eqref{prob:DIP} can be decoupled as
\begin{equation}
\textrm{REG}= \textrm{REG}^{\text{I}} + \textrm{REG}^{\text{E}}_{\text{spec}} ,
\end{equation}
which corresponds to the implicit part (i.e.,  $\textrm{REG}^{\text{I}}$) and the explicit part (i.e.,  $\textrm{REG}^{\text{E}}_{\text{spec}}$).
The subscript ``spec'' refers to spectral dimension, and is used to distinguish the spatial dimension, as will be discussed later.

The $\textrm{REG}^{\text{I}}(\bY_H)$ refers to the implicit handcrafted prior posed by the deep unfolding network itself [i.e., deep image prior (DIP) theory \cite{DIP,PRIME}], which is highly effective for spatial recovery.
Specifically, DIP induces the output image with some natural properties (e.g., self-similarity), which are demonstrated in diverse spatial restoration tasks (e.g., denoising and inpainting) \cite{DIP}.
While $\textrm{REG}^{\text{I}}$ excels at recovering spatial structures, it does not constrain the spectral dimension, potentially leading to physically unnatural spectral reconstruction.
%
Given the very limited prior knowledge about the spectral shape of the pixels in $\bY_H$, we would at most encourage the spectral smoothness in order to avoid some unnatural characteristics in the resulting solution $\widetilde{\bY}_H$.
Specifically, we define the explicit part as the spectral total-variation (TV) function to encourage spectral smoothness, i.e., $\textrm{REG}^{\text{E}}_{\text{spec}}(\bY_H)=\textrm{TV}_\textrm{spec}(\bY_H)=
\frac{1}{L(M-1)}
\sum_{\ell=1}^L
\sum_{m=1}^{M-1}
|\bY_H(m+1,\ell)-\bY_H(m,\ell)|$.
All in all, to have a lightweight network, we propose to design the deep unfolding network based on the concise criterion of
\begin{equation}\label{prob:DIP-I}
\widetilde{\bY}_H
:=
\arg\min_{\bY_H}\|\bY_S-\bD\bY_H\|_F^2+ \textrm{REG}^{\text{I}}(\bY_H),
\end{equation}
while the explicit term $\textrm{REG}^{\text{E}}_{\text{spec}}(\bY_H)$ is used as part of the loss function when training the deep unfolding network.

The architecture of the deep unfolding network is designed in the left part of Figure \ref{fig:whole_model}.
To see the design philosophy, we adopt the ADMM algorithm \cite{CVXbookCLL2016} to solve \eqref{prob:DIP-I}, followed by unfolding each ADMM closed-form solution as a deep architecture, thereby achieving an explainable network.
To this end, we reformulate \eqref{prob:DIP-I} into the ADMM form, i.e.,
\begin{equation}\label{prob:minimize}
\min_{\bY_H=\bV}
\|\bY_S-\bD\bY_H\|_F^2
+\textrm{REG}^{\text{I}}(\bV),
\end{equation}
where $\bV$ and $\bY_H$ are independent in the objective function and linearly associated in the optimization constraint, hence satisfying the standard ADMM optimization form \cite{CVXbookCLL2016}.
Then, at the $(k+1)$th iteration, the ADMM algorithm solves \eqref{prob:minimize} by iteratively updating the following two primal variables, i.e., 
\begin{align}
\bV^{k+1}
&\in\arg\min_{\bV}~\mathcal{L}_\rho(\bY_H^{k},\bV,\bU^k),
\label{eq:V-update}
\\
\bY_H^{k+1}
&\in\arg\min_{\bY_H}~\mathcal{L}_\rho(\bY_H,\bV^{k+1},\bU^k),
\label{eq:Y-update}
\end{align}
where $\mathcal{L}_\rho (\bY_H,\bV,\bU)
=
\|\bY_S-\bD\bY_H\|_F^2
+ \textrm{REG}^{\text{I}}(\bV)
+\frac{\rho}{2} \|\bY_H-\bV-\bU\|_F^2$ is the augmented Lagrangian of \eqref{prob:minimize} with $\rho>0$ being a learnable penalty parameter (cf. Figure \ref{fig:whole_model}), followed by updating the dual variable using the ADMM update rule of $\bU^{k+1}:=\bU^k-\bY_H^{k+1}+\bV^{k+1}\in\mathbb{R}^{M\times L}$ (and $k:=k+1$), until predefined stopping criterion is met.
Once the algorithm terminates at the $K$th iteration, $\widetilde{\bY}_H:=\bV^4$ will be serving as the optimal solution of \eqref{prob:minimize}.
As for how to implement \eqref{eq:V-update} and \eqref{eq:Y-update} using an explainable deep network, this can be done using the model architectures presented in Figure \ref{fig:whole_model}(c), and will be detailed later in Section \ref{sec:ID}.

For now, we discuss a more serious issue encountered when spectrally super-resolving the multi-resolution imagery, which often results in a spatially blurred solution $\widetilde{\bY}_H$ (cf. Figure \ref{fig:overview}).
The reason is that the input Sentinel-2 image $\bY_S$ contains LR 60-m bands (and some medium-resolution 20-m bands).
As spectral super-resolution can be regarded as performing computational light-splitting on $\bY_S$ \cite{PRIME}, the resultant splitting bands naturally have similar spatial resolutions as those input bands.
Therefore, the spectral-spatial duality theory, originally proved in \cite[Theorem 1]{COS2A}, proposes to solve the dilemma using a well-known technique called coupled non-negative matrix factorization (NMF) \cite{NMF,COCNMF} in the remote sensing area.
The coupled-NMF technique essentially fuses the detailed spatial information (learned from an HR reference $\widetilde{\bY}_S$) into the blurred HSI $\widetilde{\bY}_H$, thereby obtaining the desired HR HSI $\bY_H^\star$ (i.e., AVIRIS-level HSI), as illustrated in Figure \ref{fig:overview}.
The HR reference can be conveniently built by concatenating the four 10-m HR bands of ${\bY}_S$ to form the required 4-band HR MSI $\widetilde{\bY}_S$.
The above fusion strategy elegantly coincides with a widely known spatial super-resolution problem in satellite remote sensing \cite{COCNMF}, which is better studied compared to the tough spectral super-resolution problem.

To be more specific, on the basis of \cite[Theorem 1]{COS2A}, one NMF extracts spatial information from the HR MSI $\widetilde{\bY}_S$, and the other NMF extracts spectral information from the LR HSI $\widetilde{\bY}_H$.
Hence, the coupled-NMF fuses the extracted spatial and spectral information to obtain the target HR HSI $\bY_H^\star$ (cf. Figure \ref{fig:overview}).
Although the coupled-NMF can be implemented using ADMM (see, e.g., \cite{COCNMF}), the resulting convex optimization-based coupled-NMF (CO-CNMF) algorithm is not fast enough to process large-scale images \cite{COS2A}.
On the other hand, almost all the existing HSI/MSI fusion algorithms (e.g, CO-CNMF \cite{COCNMF}) require the spectral response function as part of their inputs, not user-friendly.
Therefore, we are motivated to implement the HSI/MSI fusion using a customized simple lightweight deep spatiospectral network [cf. Figure \ref{fig:whole_model}(b)], which requires only $(\widetilde{\bY}_S,\widetilde{\bY}_H)$ as its input, as will be detailed in Section \ref{sec:ID}.
As it turns out, the induced ExplainS2A algorithm (cf. Figure \ref{fig:overview}) is able to spectrally super-resolve a million-scale Sentinel-2 image $\bY_S$ into a high-standard AVIRIS-level image $\bY_H^\star$, within less than one second on a normal desktop computer (cf. Section \ref{sec: experiment}), requiring no additional input information other than $\bY_S$.
The overview of ExplainS2A is presented in Figure \ref{fig:overview}, while the detailed deep architecture is provided in Figure \ref{fig:whole_model}.
The implementation details of the proposed ExplainS2A algorithm will be collectively presented in Section \ref{sec:ID} for reproducibility.

\subsection{Implementation Details of ExplainS2A}\label{sec:ID}

Let us begin by illustrating how to implement \eqref{eq:V-update} and \eqref{eq:Y-update} using deep unfolding.
By \eqref{eq:V-update}, \eqref{eq:Y-update}, and the definition of the augmented Lagrangian $\mathcal{L}_\rho$, their closed-form solutions can be derived as
\begin{align}
\bV^{k+1}
&:=
\textrm{prox}_{\frac{1}{\rho} \textrm{REG}^{\text{I}}}(\bY_H^k-\bU^k), 
\label{prob:Z-Update}
\\
\bY_H^{k+1}
&:=
(2\bD^T\bD+\rho \bI)^{-1}\left(2\bD ^T \bY_S+\rho(\bV^{k+1}+\bU^k)\right) \notag
\\
&= 
\frac{1}{\rho}\left(\bI - \frac{2}{\rho}\bD^T
{\bm\Phi}
\bD\right)\bX^k,
\label{prob:Y_H-Update}
\end{align}
where $\textrm{prox}_{f}(\bZ)\triangleq\arg\min_{\bV} f(\bV)+\frac{1}{2}\|\bV-\bZ\|_F^2$ is the proximal operator associated with a given function $f$,
${\bm\Phi}\triangleq (\bI+\frac{2}{\rho}\bD\bD^T)^{-1} \in \mathbb{R}^{M_m\times M_m}$ is a symmetric matrix, 
and $\bX^k\triangleq 2\bD ^T \bY_S+\rho(\bV^{k+1}+\bU^k)\in\mathbb{R}^{M\times L}$ is an intermediate HSI.
The proof of \eqref{prob:Z-Update} follows from the definition of the proximal operator \cite{parikh2014proximal}, while the proof of \eqref{prob:Y_H-Update} follows from the matrix inversion lemma (i.e., Woodbury matrix inversion identity) \cite{CVXbookCLL2016}.
The proof details are omitted for conciseness.

To implement \eqref{prob:Z-Update} and \eqref{prob:Y_H-Update} using an explainable deep network, we discuss their physical meaning.
 For \eqref{prob:Z-Update}, the proximal operator can be interpreted as a denoiser within the maximum a posteriori (MAP) sense \cite{meinhardt2017learning}.
In addition to traditional denoisers, such as non-local means (NLmeans) \cite{NLM} and block-matching and 3-D filtering (BM3D) \cite{BM3D}, recent advancements have demonstrated that denoising convolutional neural network (DnCNN) \cite{DNCNN} can effectively serve as learnable proximal mappings to capture more sophisticated image priors \cite{meinhardt2017learning}.
Therefore, the noisy image ``$\bY_H^k-\bU^k$'' is formed at each stage $k$, and then fed into Model 1 for denoising, as illustrated in Figure \ref{fig:whole_model}(a).
%
%
%
Considering that a network with larger $K$ converges slowly during the training phase, violating the trend of developing green artificial intelligence \cite{aggarwal2025reducing}, we empirically select the model-order as $K:=4$ to strike a balance between the convergence rate and super-resolution performances.
Since Model 1 is for denoising, it is natural to adopt the residual-in-residual architecture that is strong for the denoising task \cite{yao2022dense}, as detailed in Figure \ref{fig:whole_model}(c), where ``k$3$n$172$'' denotes the convolution block using $3\times 3$ kernel over $172$ feature channels \cite{CODE}.
The residual-in-residual network itself serves as an implicit yet effective deep prior according to the DIP theory \cite{DIP}, as discussed in Section \ref{sec:algodesign}.
%
%
On the other hand, in order to unfold \eqref{prob:Y_H-Update}, the image $\bX^k$ is also formed at each stage $k$ [cf. Figure \ref{fig:whole_model}(a)], according to its definition in \eqref{prob:Y_H-Update}, and then fed into Model 2 [cf. Figure \ref{fig:whole_model}(c)] that implements the operator of $\frac{1}{\rho}\left(\bI - \frac{2}{\rho}\bD^T{\bm\Phi}\bD\right)$ in \eqref{prob:Y_H-Update}.
There are two tracks in this operator.
The first track ``$\bI$'' corresponds to the identity shortcut connection in Model 2 [cf. Figure \ref{fig:whole_model}(c)].
The second track ``$\bD^T{\bm\Phi}\bD$'' sequentially implements $\bD$ (i.e., downsampling), ${\bm\Phi}$ [i.e., fully connected (FC) layer] and $\bD^T$ (i.e., upsampling) from right to left [cf. Figure \ref{fig:whole_model}(c)].
Note that any matrix $\bm\Phi$ can be implemented using FC layer.
We implement it using a symmetric FC (SFC) layer as $\bm\Phi$ defined in \eqref{prob:Y_H-Update} is clearly a symmetric matrix.
The scaling factors $(-2/\rho,1/\rho)$ in \eqref{prob:Y_H-Update} are also considered in Model 2, as illustrated in Figure \ref{fig:whole_model}(c).

Once we obtain $\widetilde{\bY}_H$ from the Model 1 and Model 2 designed above [cf. Figures \ref{fig:whole_model}(a) and \ref{fig:whole_model}(c)], the remaining task is to implement the HSI/MSI fusion for further obtaining the target $\bY_H^\star$.
As discussed in Section \ref{sec:algodesign}, the fusion task can be elegantly done using convex optimization \cite{COS2A}, while, considering the computational efficiency, we will propose a simple lightweight deep network [cf. Figure \ref{fig:whole_model}(b)] to fuse $\widetilde{\bY}_H$ and $\widetilde{\bY}_S$.
The design philosophy will be learning spectral attention from $\widetilde{\bY}_H$ [cf. Model 3 in Figure \ref{fig:whole_model}(c)] and spatial attention from $\widetilde{\bY}_S$ [cf. Model 4 in Figure \ref{fig:whole_model}(c)], and use the learned information to assist learning the residual $\bY_\textrm{res}$ [cf. Model 5 in Figure \ref{fig:whole_model}(c)] for reconstructing the target image $\bY_H^\star$.
The physical meaning of these fusion models will be designed and explained next.

Let $\bm 1_z$ denote the $z$-dimensional all-one vector, $\bI_z$ denote the $z\times z$ identity matrix, and $\textrm{Diag}(\bw)$ denote the diagonal matrix with its $i$th diagonal entry being $[\bw]_i$.
The intermediate HSI $\widetilde{\bY}_H$ is first spatially downsampled as $\widetilde{\bY}_H^{\downarrow} \in \mathbb{R}^{M\times (L/4)}$ to reduce the number of parameters, and then fed into Model 3 to learn the spectral attention vector, i.e.,
\begin{equation}\label{eq:wspec}
\bw_{\text{spec}} = \text{sigmoid}\Big(\text{FC}\!\left( 
\bv_{\text{spec}}
\right)\Big)\in\mathbb{R}^M,
\end{equation}
in order to emphasize the informative spectral components [cf. Figure \ref{fig:whole_model}(b)].
Here, $\bv_{\text{spec}}\triangleq \frac{4}{L}\widetilde{\bY}_H^{\downarrow} {\bm 1}_{L/4}$ is the average pooling result [cf. Figure \ref{fig:whole_model}(c)].
In parallel, $\widetilde{\bY}_S$ is undergoing Model 4 to learn the spatial attention vector, i.e.,
\begin{equation}\label{eq:wspat}
\bw_{\text{spat}} = \text{sigmoid}\Big(\text{Conv.}\!\left(\bv_{\text{spat}}\right)\Big)
\in \mathbb{R}^{L},   
\end{equation}
where $\bv_{\text{spat}}\triangleq \frac{1}{4} {\bm 1}_4^T \widetilde{\bY}_S$ is the channelwise mean of the 4 HR Sentinel-2 bands, and $\text{Conv.}\!\left(\cdot\right)$ is a single-channel $5 \times 5$ convolution layer [cf. Figure \ref{fig:whole_model}(c)].
The spatial attention vector $\bw_{\text{spat}}$ effectively captures the relative significance of spatial features.
Then, a spatiospectral-emphasized HSI feature is computed as 
\begin{equation}
\label{prob:spectral enhance}
\widetilde{\bY}_{\text{spec-spat}}
= 
\textrm{Diag}(\bw_\textrm{spec}) 
\widetilde{\bY}_H^{\uparrow} 
\textrm{Diag}(\bw_\textrm{spat}), 
\end{equation}
where $\widetilde{\bY}_H^{\uparrow} \triangleq \widetilde{\bY}_H^{\downarrow} \left( \bI_{L/4}\otimes {\bm 1}_4^T \right)\in\mathbb{R}^{M\times L}$ that is upsampled to have the same dimension as the target HSI [cf. Figure \ref{fig:whole_model}(b)], and the notation $\otimes$ denotes the Kronecker product.

Finally, the emphasized $M$-band HSI feature $\widetilde{\bY}_{\text{spec-spat}}$ is concatenated with the $4$-band HR feature $\widetilde{\bY}_S$, and then convoluted and activated to obtain the fused $(M+4)$-channel feature maps, denoted as $\bZ\in\mathbb{R}^{(M+4)\times L}$, which are subsequently processed by Model 5 for learning the residual $\bY_\textrm{res}$ as illustrated in Figure \ref{fig:whole_model}(b).
Model 5 progressively refines the feature maps in $\bZ$ by simple residual networks [cf. Figure \ref{fig:whole_model}(c)], and eventually obtains the residual information $\bY_\textrm{res}$ (the missing details) to be added back to the upsampled HSI $\widetilde{\bY}_H^{\uparrow}$, thereby yielding the target AVIRIS-level HSI $\bY_H^\star$ [cf. Figure \ref{fig:whole_model}(b)].
Besides the implicit network prior, considering that the deep fusion aims to enhance the spatial quality, we explicitly regularize the network via a spatial TV loss function $\textrm{REG}^{\text{E}}_{\text{spat}}(\bY_H)=\textrm{TV}_\textrm{spat}(\bY_H)$ (i.e., the normalized $\ell_1$-norm of the spatial gradients) \cite{CDADMM}, which is defined in a similar way as $\textrm{REG}^{\text{E}}_{\text{spec}}(\bY_H)$ (cf. Section \ref{sec:algodesign}).
This completes the design of the lightweight yet highly effective fusion network, which will be trained together with the unfolding network [i.e., Figure \ref{fig:whole_model}(a)] in an end-to-end manner based on the ADAM optimizer \cite{kingma2014adam}.

As it turns out, the induced ExplainS2A algorithm is able to solve the highly ill-posed spectral super-resolution problem for fast yielding high-fidelity AVIRIS-level HSI (cf. Section \ref{sec: experiment}).
This allows historical Sentinel-2 data to be transformed into their counterpart high-standard AVIRIS data, and greatly facilitates the downstream tasks that rely on the strong identifiability of HSI (e.g., blind source separation) as will be demonstrated in Section \ref{sec:CaseStudy}.

\subsection{Extension of ExplainS2A to General Sensor Pairs}\label{sec: general}

While this paper primarily demonstrates the transformation from Sentinel-2 data to AVIRIS image as a representative study, it is crucial to emphasize that the proposed ExplainS2A is a generic framework based on the spectral-spatial duality theory \cite[Theorem 1]{COS2A}.
The underlying mathematical formulation (cf. Section \ref{sec:algodesign} and Section \ref{sec:ID}) is independent of the sensor hardware, so that ExplainS2A can be extensively applied to various multispectral/hyperspectral sensor pairs.
We will describe the adaptation strategies for two categories of multispectral sensors to illustrate the generalizability of our framework.

First, for multispectral sensors that contain multi-resolution bands (i.e., with some LR bands similar to Sentinel-2), the adaptation only requires the preparation of corresponding training pairs and the adjustment of the input/output channel dimensions to align with the multispectral/hyperspectral sensor specifications.
%
Second, for multispectral sensors that have single-resolution bands, the implementation is even more streamlined. 
Specifically, in the deep fusion stage, the band selection step can be omitted (namely, $\bY_S$ can be directly used as the input instead of the selected subset $\widetilde{\bY}_S$).
Furthermore, given the absence of the multi-resolution issue, the super-resolved $\widetilde{\bY}_H$ inherently possesses HR spatial information, for which the fusion network may concentrate more on learning the spectral reconstruction part via the spectral attention mechanism (cf. Figure \ref{fig:whole_model}).

\section{Experimental Results and Discussion}\label{sec: experiment}

In this section, we demonstrate the effectiveness, computational efficiency, and real-world application of the proposed ExplainS2A algorithm.
In Section \ref{sec:experimental setting}, we present the experimental settings, including the experimental design and dataset description.
In Section \ref{sec:Results}, we conduct comprehensive quantitative and qualitative analysis for understanding the strength of the method.
%
%
In Section \ref{sec:TimeAnalysis}, we investigate the complexity and scalability of ExplainS2A, and demonstrate the superiority of ExplainS2A when processing large-scale images.
In Section \ref{sec:discussions}, we present a comprehensive analysis of ExplainS2A, comprising an ablation study, challenging condition analysis, as well as a systematic comparison with COS2A, which is the only existing method that addresses the same challenging multi-resolution scenario.
In Section \ref{sec:CaseStudy}, we conduct a case study to show that performing blind source separation on ExplainS2A-reconstructed HSI does yield much better and more explainable results than performing it on the original Sentinel-2 data.

%
%
%
%

\subsection{Experimental Setting}\label{sec:experimental setting}

Since there is only one prior work considering the target problem (i.e., conversion of Sentinel-2 data to AVIRIS data) \cite{COS2A}, we follow the experimental protocol designed therein for fair performance evaluation.
To ensure that the reconstructed HSIs maintain high spectral fidelity despite the different spectral coverage of the two sensors, experimental data are collected from a spatiotemporally diverse dataset \cite[Table 1]{DCSN}.
Specifically, there are 646 AVIRIS/Sentinel-2 data pairs ($\bY_H,\bY_S$) with ground-truth (GT) $\bY_H$ for quantitative comparison experiments, and they are divided into training/validation/testing data based on the ratio of 15:1:1. 
Each AVIRIS hyperspectral data \cite{AVIRISdata} has 256 $\times$ 256 pixels.
Consistent with prior studies \cite{HyperQUEEN, dehazing, DCSN}, we excluded spectral bands severely degraded by water vapor absorption (i.e., bands 1–10, 104–116, 152–170, and 215–224).
Consequently, a subset of 172 usable bands is retained from the original 224 bands, covering wavelengths ranging from 400 to 2500 nm.
%
A total of 20000 spatially overlapped 64$\times$64 patches are cropped during the training/validation procedure.
%
%
%
%
Other details about the dataset customized for the target spectral super-resolution problem are available from \cite[Section III.A]{COS2A}.

\begin{figure}[t]
    \centerline{\includegraphics[width=0.5\textwidth]{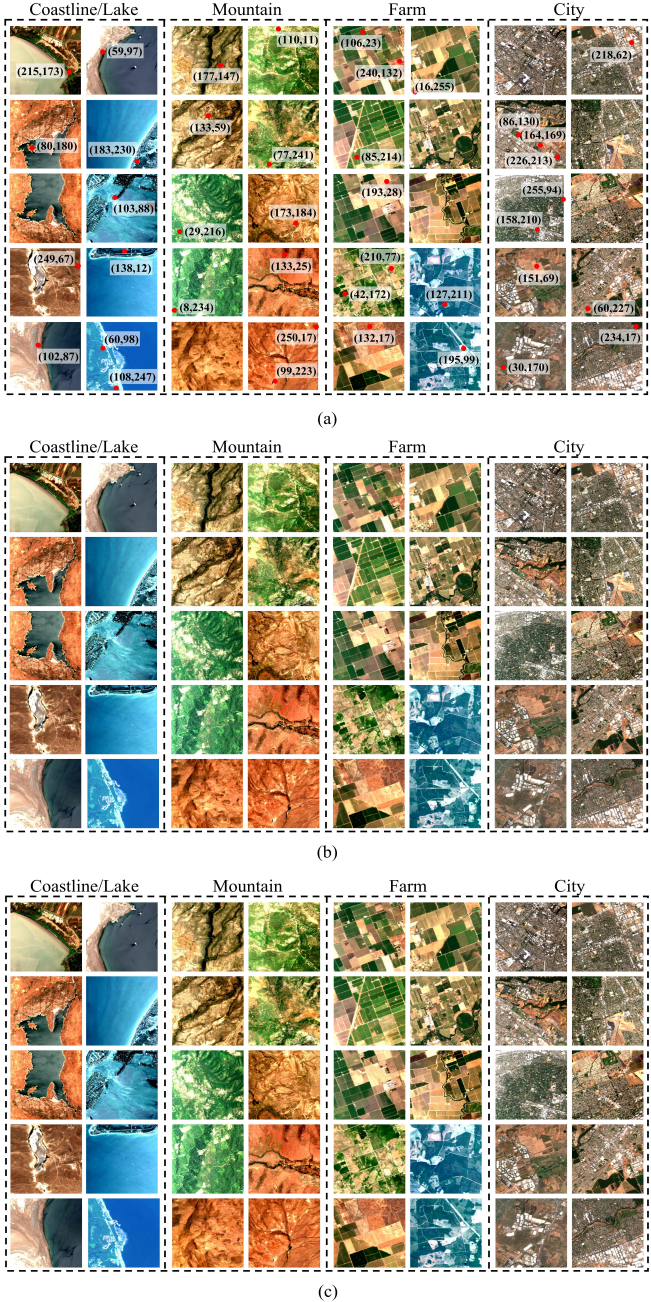}}
    \caption{(a) The reference AVIRIS data $\bY_H$ over four representative land types, where 10 representative pixels (i.e., spectrally distinct pixels, marked by red dots) are selected from each land type for evaluating the spectral fidelity of the reconstructed hyperspectral pixels (cf. Figure \ref{fig: spectral}).
    %
    The reconstructed AVIRIS-level data $\bY_H^\star$ by (b) SSU-Net and (c) ExplainS2A.
    The images displayed here are represented using the true-color composition [B23(r)-B12(g)-B5(b)].
    }\label{fig: 40}
\end{figure}

\begin{figure}[t]
    \centerline{\includegraphics[width=0.5\textwidth]{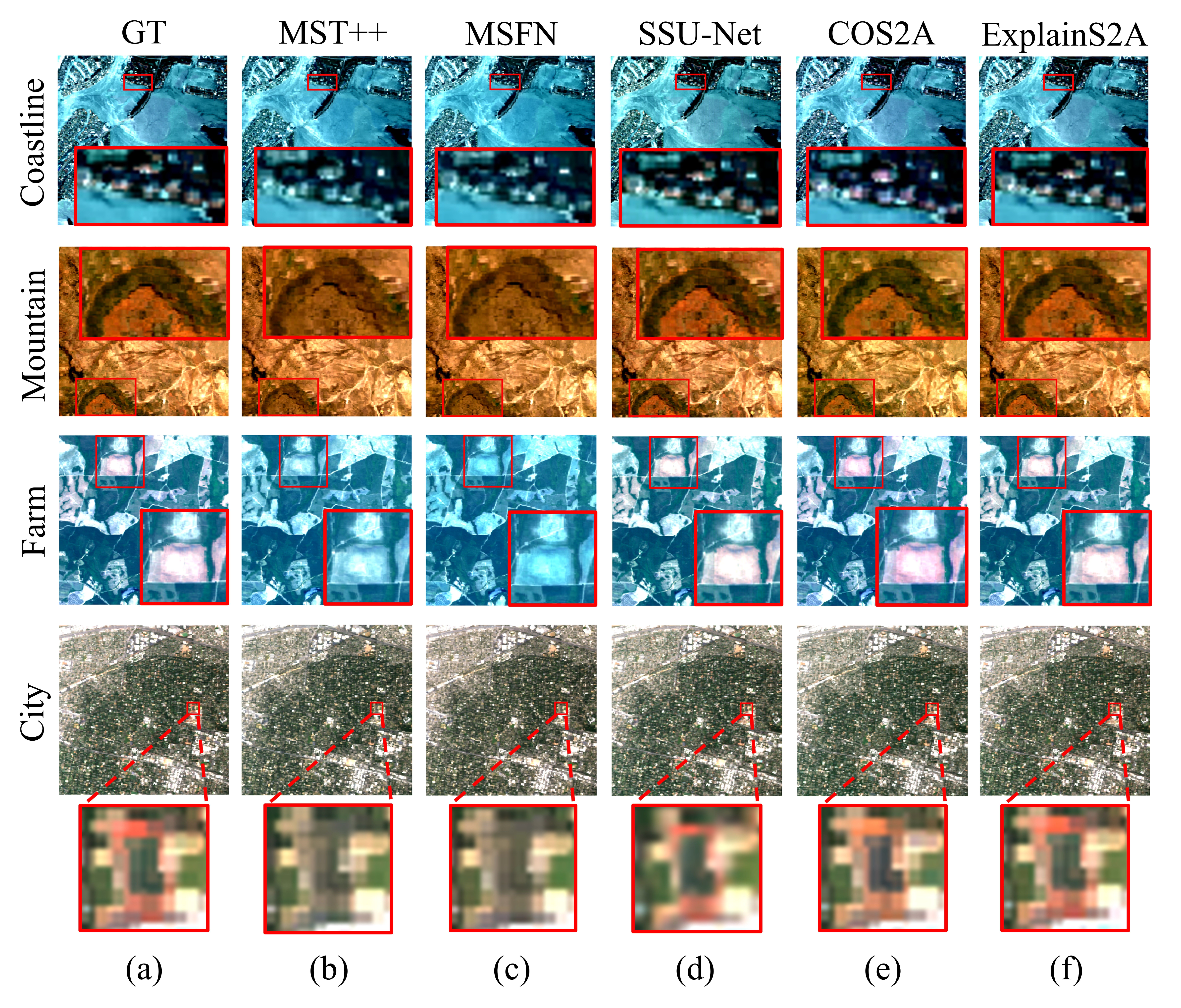}}
    \caption{(a) The reference GT over four representative land types. 
    The spatial reconstruction results obtained by (b) MST++, (c) MSFN, (d) SSU-Net, (e) COS2A, and (f) the proposed ExplainS2A.
    All images are shown in true-color composition [B23(R)–B12(G)–B5(B)] over the four regions of interest (ROIs).
    We further investigate the fine details of the city landscape as it has been reported to be the most intricate terrain \cite[Figure 9]{COS2A}, making it a suitable arena to demonstrate the spatial super-resolution results.
    }\label{fig: 40_results}
\end{figure}

\begin{figure}[t]
    \centerline{\includegraphics[width=0.5\textwidth]{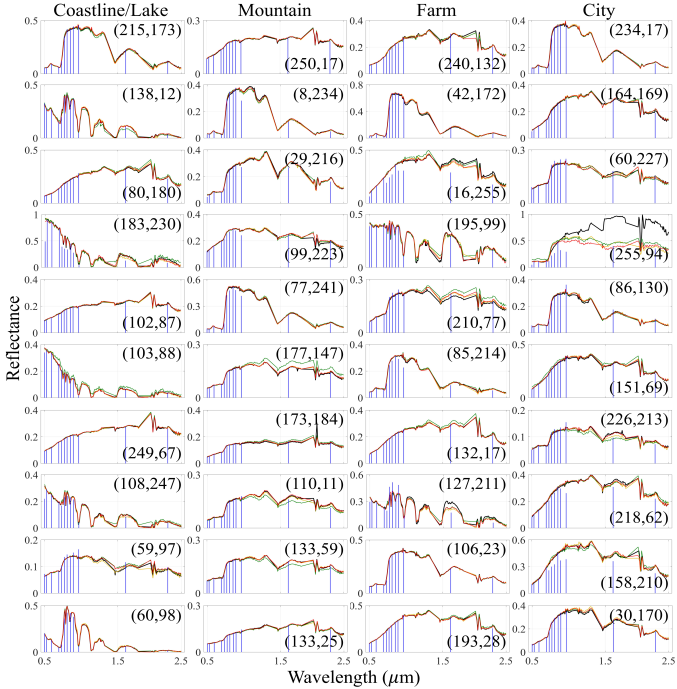}}
    \caption{Spectral reconstruction results of SSU-Net, COS2A, and the proposed ExplainS2A algorithm, displayed in the target wavelength range (i.e., 400 to 2500 nm), over the representative pixels from each land type (cf. Figure \ref{fig: 40}).
    The coordinates of these pixels are marked in Figure \ref{fig: 40}(a).
    The corresponding 12-band Sentinel-2 pixels are displayed as blue pulses.
    The hyperspectral pixels reconstructed by SSU-Net, COS2A, and ExplainS2A are displayed as yellow, green, and red curves, respectively, while the real AVIRIS pixels (i.e., GT pixels) are displayed as black curves.
    }\label{fig: spectral}
\end{figure}

Accordingly, the dataset is used to train the proposed ExplainS2A algorithm using the following loss function, i.e.,
\[
\|\widetilde{\bY}_H-\bY_H\|_1
+\|\bY_H^\star-\bY_H\|_1
+\lambda[\textrm{REG}^{\text{E}}_{\text{spec}}(\widetilde{\bY}_H) 
+ \textrm{REG}^{\text{E}}_{\text{spat}}(\bY_H^\star)], 
\]
where $\bY_H$ is the GT from the dataset, 
$\widetilde{\bY}_H$ is the output of the deep unfolding network [cf. Figure \ref{fig:whole_model}(a)],
$\bY_H^\star$ is the output of the ExplainS2A [cf. Figure \ref{fig:whole_model}(b)], 
the regularization parameter is empirically set as $\lambda:=10^{-4}$,
and $\textrm{REG}^{\text{E}}_{\text{spec}}(\widetilde{\bY}_H)$ and $\textrm{REG}^{\text{E}}_{\text{spat}}(\bY_H^\star)$ are defined in Section \ref{sec:algodesign} and Section \ref{sec:ID}, respectively.
The design philosophy is natural.
The first two terms are to force the intermediate result and the final result to be as close to GT as possible, and use the $\ell_1$-norm to achieve outlier-robust data fitting \cite{CODE}.
The last two terms are the explicit regularizers, which explicitly encourage spectral and spatial smoothness on $\widetilde{\bY}_H$ and $\bY_H^\star$, respectively, as discussed in Section \ref{sec:thm}.
To enhance spectral fidelity, we recall that a spectral attention mechanism is additionally integrated into the deep fusion network to capture and prioritize the relative significance of spectral features (cf. Section \ref{sec:ID}).
%
The initial learning rate is set as 0.0001, followed by the multistep learning rate scheduler, where the learning rate is reduced by a factor of 0.5 at epochs 30, 60, and 90, to facilitate smooth and stable convergence, and the overall ExplainS2A algorithm is trained in an end-to-end manner using the ADAM optimizer \cite{kingma2014adam}.
%

As for the baseline preparation, to the best of our knowledge, COS2A \cite{COS2A} is the only existing work specifically tailored for transforming the multi-resolution Sentinel-2 data to AVIRIS image across the visible to NIR bands; thus, it is naturally selected as the key baseline.
Given the scarcity of direct competitors, we extend our comparison to include the seminal MST++ \cite{MST++} and the recent MSFN \cite{MSFN}, which were originally designed for pure spectral super-resolution within the visible bands.
To bridge the gap caused by different problem settings and enable a meaningful comparison, we follow the modification proposed in \cite{COS2A} when training these methods using the divide-and-conquer strategy \cite{DACalgo}.
This modification maximizes the preservation of their original architectures while enabling them to handle 172-band spectral outputs.
%
%
For a more comprehensive comparison, we directly include SSU-Net as a baseline \cite{SSU-Net}, which can handle the reconstruction of more than one hundred bands.
We only modify the number of input/output channels to meet the target mission, thereby retaining the original architecture design of SSU-Net.
%
All the experiments are conducted under MATLAB R2025a, Python 3.12.3, and PyTorch 2.8.0 on a computer equipped with an Intel Core i9-10900K CPU with 3.70-GHz speed, and 62-GB of system RAM, and an NVIDIA GeForce RTX 3090 GPU.

\subsection{Qualitative and Quantitative Analysis}\label{sec:Results} 

To qualitatively evaluate the similarity between the ExplainS2A-generated $\bY_H^\star$ and the real AVIRIS image $\bY_H$, we use the testing dataset adopted in the COS2A work \cite{COS2A}.
For self-contained purposes, the 40 testing regions of interest (ROIs), uniformly sampled from each land type, are displayed in Figure \ref{fig: 40}(a).
Ten spectrally distinct hyperspectral pixels are also selected as the representatives for each land type by following the testing protocol designed in \cite{COS2A}, and they are marked by red dots in Figure \ref{fig: 40}(a) for the subsequent qualitative evaluation of the spectral fidelity (cf. Figure \ref{fig: spectral}).
The reconstruction results from MST++, MSFN, and COS2A can be found from \cite[Figure 5(a)]{COS2A}, \cite[Figure 5(b)]{COS2A}, and \cite[Figure 5(c)]{COS2A}, respectively, and they are not displayed here for conciseness.
Then, the HSIs generated by SSU-Net and the proposed ExplainS2A algorithm are summarized in Figure \ref{fig: 40}(b) and Figure \ref{fig: 40}(c), respectively.
%
For ease of discussion, we visualize the spatial reconstruction results of all methods using all four land types, as shown in Figure \ref{fig: 40_results}.
One can see that both the seminal MST++ method \cite{MST++} and the very recent MSFN method \cite{MSFN} suffer from serious color distortion.
This phenomenon highlights the inadequacy of these baselines designed for single-resolution, visible-spectrum images when applied to the multi-resolution, full-spectrum task.
Therefore, it is motivated to use the spectral-spatial duality \cite[Theorem 1]{COS2A} to effectively solve such a highly complicated inverse problem.
Under the spectral-spatial duality, COS2A \cite{COS2A} only exhibits slight color shifts; for instance, the results appear slightly reddish in the coastline and farm scenes [cf. Figure \ref{fig: 40_results}(e)].
Furthermore, the reconstructed HSIs from SSU-Net and the proposed ExplainS2A are more consistent with GT in the general land types such as coastline, mountain, and farm scenes [cf. Figures \ref{fig: 40_results}(d) and \ref{fig: 40_results}(f)].
However, in the most intricate city scene, SSU-Net yields comparatively smoother reconstructions, resulting in the loss of fine spatial details [cf. Figure \ref{fig: 40_results}(d)].
For example, within the red boundary area of the zoomed-in frame (particularly in the upper and left regions), the width is inconsistent with GT [cf. Figure \ref{fig: 40_results}(d)].
Also, SSU-Net produces an L-shaped pattern in the central region, whereas COS2A and ExplainS2A correctly preserve the Z-shaped structure [cf. Figures \ref{fig: 40_results}(e) and \ref{fig: 40_results}(f)], echoing the GT reference.
Overall, ExplainS2A demonstrates the best spatial reconstruction performance across all land types.

\begin{table}[t!]
\footnotesize
\centering
\caption{Quantitative evaluation of the spectral super-resolution methods, including MST++ \cite{MST++}, MSFN \cite{MSFN}, SSU-Net \cite{SSU-Net}, COS2A \cite{COS2A}, and the proposed ExplainS2A algorithm.
Each boldfaced number represents the best performance of a land type for a specific performance index (e.g., the largest PSNR/SSIM, and the smallest SAM/RMSE/Time).}
\label{tab: quan}
\setlength{\tabcolsep}{1mm}
\renewcommand{\arraystretch}{1.2}
\scalebox{0.92}{
\begin{tabular}{c|c|ccccc}
\hline
& Methods & PSNR $\!(\uparrow)$ & SAM $\!(\downarrow)$ & RMSE $\!(\downarrow)$ & SSIM $\!(\uparrow)$ & Time (sec.) \\
\hline
\multirow{5}{*}{\makecell{Coastline/\\Lake}}
& MST++ & 27.6972 & 8.4138 & 0.0264 & 0.9285 & 0.0665\\
\cdashline{2-7}
& MSFN & 28.5163 & 7.5827 & 0.0257 & 0.9464 & 1.3051 \\
\cdashline{2-7}
& SSU-Net & 37.5741 &  2.1833 & 0.0063 &  0.9825 & 0.0793\\
\cdashline{2-7}
& COS2A & 35.8897 & 2.4539 & 0.0080 & 0.9704 & 15.3353 \\
\cdashline{2-7}
& ExplainS2A & {\bf39.1497} & {\bf1.5609} & {\bf0.0052} & {\bf0.9908} & {\bf0.0475} \\
\hline
\multirow{5}{*}{\makecell{Mountain}}
& MST++ & 28.2909 & 4.9266 & 0.0180 & 0.9317 & 0.0659 \\
\cdashline{2-7}
& MSFN & 29.2526 & 4.0895 & 0.0148 & 0.9427 & 1.3025 \\
\cdashline{2-7}
& SSU-Net & 37.6097 &  1.2176 &  0.0050 & 0.9906 & 0.0766\\
\cdashline{2-7}
& COS2A & 33.2310 & 2.0312 & 0.0092 & 0.9512 & 14.5959\\
\cdashline{2-7}
& ExplainS2A & {\bf38.1645} & {\bf1.0853} & {\bf0.0047} & {\bf0.9916} & {\bf0.0476} \\
\hline
\multirow{5}{*}{\makecell{Farm}}
& MST++ & 27.0976 & 9.0505 & 0.0427 & 0.8896 & 0.0654 \\
\cdashline{2-7}
& MSFN & 28.0107 & 8.9927 & 0.0411 & 0.9034 & 1.2987 \\
\cdashline{2-7}
&  SSU-Net & 39.0454& 1.6674 &  0.0079 & 0.9774 &  0.0762\\
\cdashline{2-7}
& COS2A & 35.0685 & 2.1460 & 0.0113 & 0.9527 & 13.9210 \\
\cdashline{2-7}
& ExplainS2A & {\bf40.5761} & {\bf1.2211} & {\bf0.0061} & {\bf0.9884} & {\bf0.0473} \\
\hline
\multirow{5}{*}{\makecell{City}}
& MST++ & 28.0807 & 8.4681 & 0.0408 & 0.8230 & 0.0676 \\
\cdashline{2-7}
& MSFN & 28.2156 & 6.2153 & 0.0341 & 0.8453 & 1.3007 \\
\cdashline{2-7}
& SSU-Net & 39.4428 & 2.0292 &  0.0108 & 0.9787 & 0.0781\\
\cdashline{2-7}
& COS2A & 35.8443 & 3.3838 & 0.0175 & 0.9209 & 14.6907 \\
\cdashline{2-7}
& ExplainS2A & {\bf39.7959} & {\bf1.9323} & {\bf0.0103} & {\bf0.9794} & {\bf0.0474} \\
\hline
\hline
\multirow{5}{*}{\makecell{Average}}
& MST++ & 27.7916 & 7.7147 & 0.0320 & 0.8932 &  0.0664\\
\cdashline{2-7}
& MSFN & 28.4988 & 6.7201 & 0.0289 & 0.9095 & 1.3018 \\
\cdashline{2-7}
&  SSU-Net & 38.4180 &  1.7744 & 0.0075 &  0.9823 & 0.0776\\
\cdashline{2-7}
& COS2A & 35.0084 & 2.5037 & 0.0115 & 0.9488 & 14.6357 \\
\cdashline{2-7}
& ExplainS2A & {\bf39.4216} & {\bf1.4499} & {\bf0.0066} & {\bf0.9876} & {\bf0.0475} \\
\hline
\end{tabular}}
\end{table}

Although SSU-Net, COS2A, and ExplainS2A yield significant spatial reconstruction results (cf. Figure \ref{fig: 40_results}), we will demonstrate the clear superiority of ExplainS2A, in terms of spectral reconstruction fidelity (cf. Figure \ref{fig: spectral}), quantitative evaluation results (cf. Table \ref{tab: quan}), and the computational efficiency when processing large-scale images (cf. Figure \ref{fig:linear}).
Let us begin with the spectral reconstruction analysis next.

In Figure \ref{fig: spectral}, for those representative GT reference pixels (displayed as black curves), the SSU-Net-reconstructed, the COS2A-reconstructed, and the ExplainS2A-reconstructed hyperspectral pixels are displayed as yellow, green, and red curves, respectively.
The blue pulses in Figure \ref{fig: spectral} are Sentinel-2 pixels.
The coordinates of the true/reconstructed pixels displayed in Figure \ref{fig: spectral} have the corresponding locations marked in Figure \ref{fig: 40}(a).
Those ExplainS2A-reconstructed pixels show better resemblance to the GT ones.
For example, at the (183,230)th and (108,247)th pixels of the coastline land type, ExplainS2A yields more precise spectral shapes in the wavelength range of 1.8-2.5 $\mu$m (cf. Figure \ref{fig: spectral}).
In the mountain and farm areas, both ExplainS2A and SSU-Net demonstrate high consistency in preserving spectral shapes at (177,147)th and (132,17)th pixels, respectively, while this is not achieved by COS2A (cf. Figure \ref{fig: spectral}).
%
%
For the city scene, we observe significant improvements of ExplainS2A at specific key pixels, for example, at the (226,213)th pixel in the wavelength range of 1.8-2.5 $\mu$m, and at the (158,210)th pixel across 1.0-1.3 $\mu$m, which is consistent with the visual advantages highlighted in Figure \ref{fig: 40_results}.
%
Overall, Figure \ref{fig: spectral} demonstrates the outstanding spectral reconstruction capability of ExplainS2A, which is able to reconstruct sharp oscillation patterns of real AVIRIS pixels in the NIR wavelength region.
The superiority of ExplainS2A will be further verified by the spectral angle mapper (SAM) and root mean squared error (RMSE) indices (cf. Table \ref{tab: quan}).

To have a fair and more convincing comparison, we conduct the quantitative evaluation of the spectral super-resolution methods, as summarized in Table \ref{tab: quan}.
Specifically, to compare the similarity between the reconstructed HSI $\bY_H^\star$ and the GT $\bY_H$, we adopt commonly used quantitative metrics, including the peak signal-to-noise ratio (PSNR), SAM, RMSE, and structural similarity (SSIM) \cite{CODEIF,SISHY}.
Given that accurate spectral recovery is the core objective of spectral super-resolution, SAM and RMSE are considered the most directly relevant metrics.
While SAM captures the spectral shape similarity, RMSE serves as a global index assessing the reconstruction accuracy in terms of element-wise magnitude.
The testing results averaged over the testing data are summarized in Table \ref{tab: quan} for the investigated methods, including MST++ \cite{MST++}, MSFN \cite{MSFN}, SSU-Net \cite{SSU-Net}, COS2A \cite{COS2A}, and the proposed ExplainS2A algorithm.
One can see that even for the seminal MST++ and the very recent MSFN, the SAM errors are often higher than 6 degrees, and the RMSE values are generally higher than 0.02, mainly because they were not developed for the target challenging problem.
%
Note that existing spectral super-resolution methods mostly target the CAVE-level 3-to-31 band reconstruction, making them far from being sufficient to address the 12-to-172 case encountered when transforming Sentinel-2 data to AVIRIS data.
COS2A is the first method that achieves the above multi-resolution 12-to-172 band reconstruction mission \cite{COS2A}, so it remarkably yields the averaged SAM lower than 3 degrees and RMSE lower than 0.02 for the first time.
%
For SSU-Net, it yields competitive results with an averaged SAM below 1.8 degrees and RMSE below 0.01, owing to its capability to process over 100 spectral bands.
Nevertheless, ExplainS2A goes a step further to reduce the SAM and the RMSE, and upgrade the PSNR and SSIM (cf. Table \ref{tab: quan}), achieving the best results across all metrics.
Regarding computational efficiency, the lightweight architecture of ExplainS2A enables a significantly shorter inference time compared to COS2A (due to the use of coupled-NMF), and it even outperforms other deep learning-based peer methods (cf. Table \ref{tab: quan}). 
In summary, the proposed ExplainS2A achieves superior reconstruction while maintaining the fastest computational speed, making ExplainS2A highly practical and effective for real-world applications.

\begin{figure}[t]
    \centerline{\includegraphics[width=0.5\textwidth]{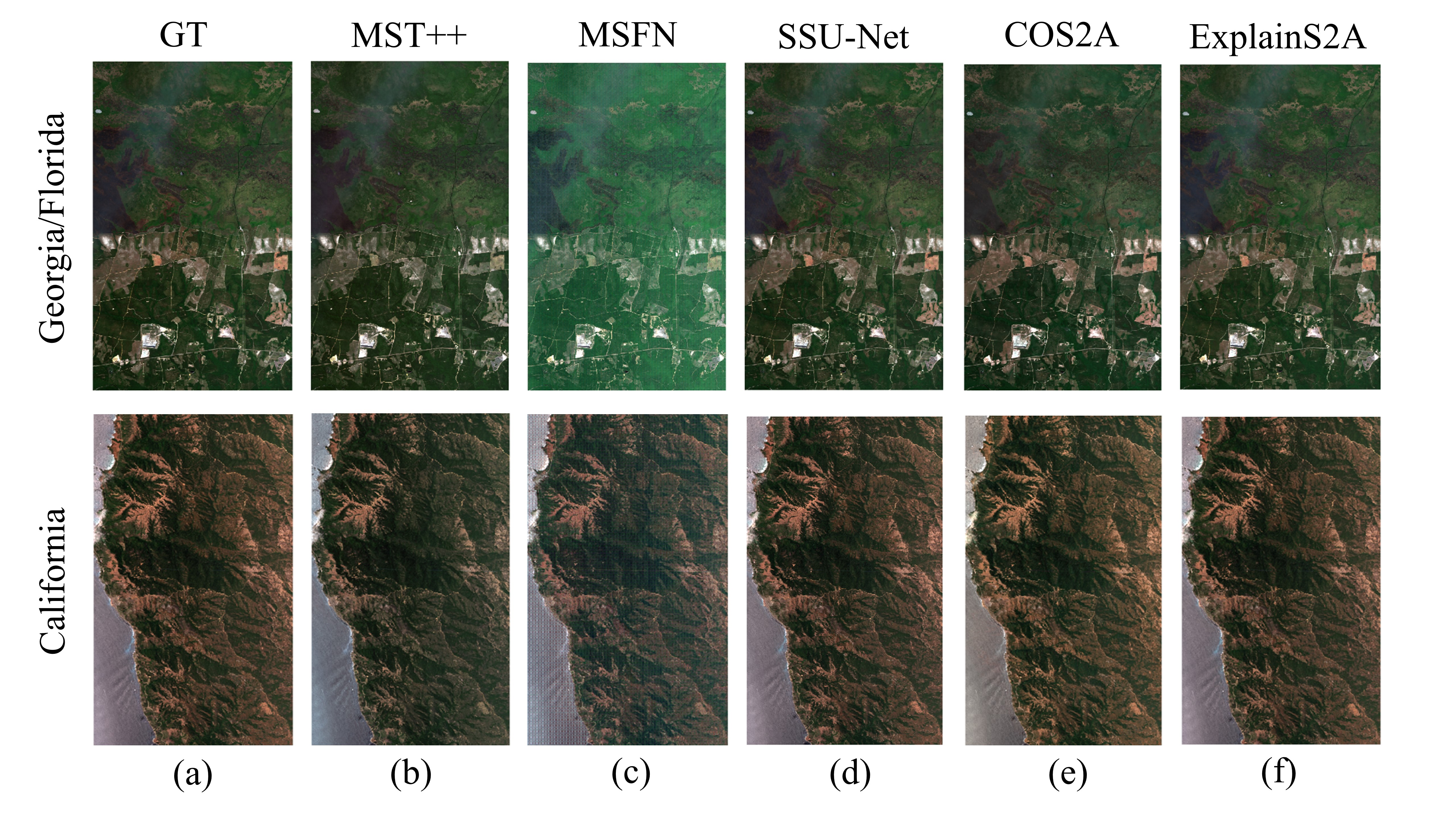}}
    \caption{(a) Testing ROIs from the GA/FL border (1090 $\times$ 1798 pixels) and the CA region (1080 $\times$ 1812 pixels), and the HSIs reconstructed by (b) MST++, (c) MSFN, (d) SSU-Net, (e) COS2A, and (f) the proposed ExplainS2A algorithm.
    }\label{fig:Figure6}
\end{figure}

\begin{figure}[t]
    \centerline{\includegraphics[width=0.5\textwidth]{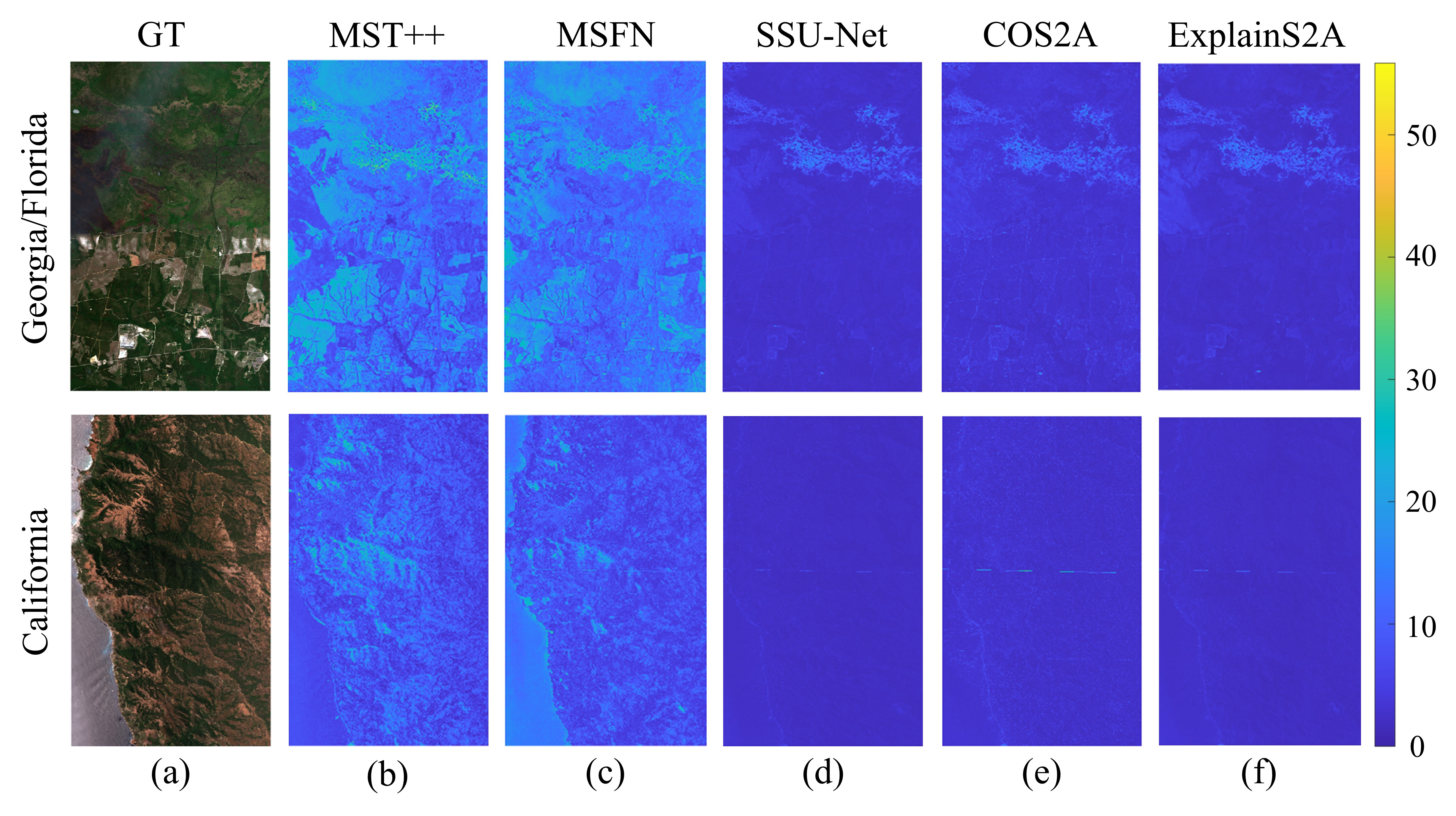}}
    \caption{(a) Testing ROIs (same as Figure \ref{fig:Figure6}), and the SAM error maps for the HSIs reconstructed by (b) MST++, (c) MSFN, (d) SSU-Net, (e) COS2A, and (f) the proposed ExplainS2A algorithm.
    The SAM value represents the spectral angle error between each reconstructed pixel in $\bY_H^\star$ and its corresponding reference pixel in the real AVIRIS image $\bY_H$.
    %
    The averaged inference times for MST++, MSFN, SSU-Net, COS2A, and ExplainS2A are approximately 2.1 sec., 1.2 minutes, 1.9 sec., 4 hours, and 1.6 sec., respectively, highlighting the superior computational efficiency of ExplainS2A compared with the time-consuming COS2A (due to the use of coupled-NMF) and even with the other deep learning-based peer methods.
    }\label{fig:Figure7}
\end{figure}

\begin{figure}[t]
\centering
\includegraphics[width=0.8\linewidth]{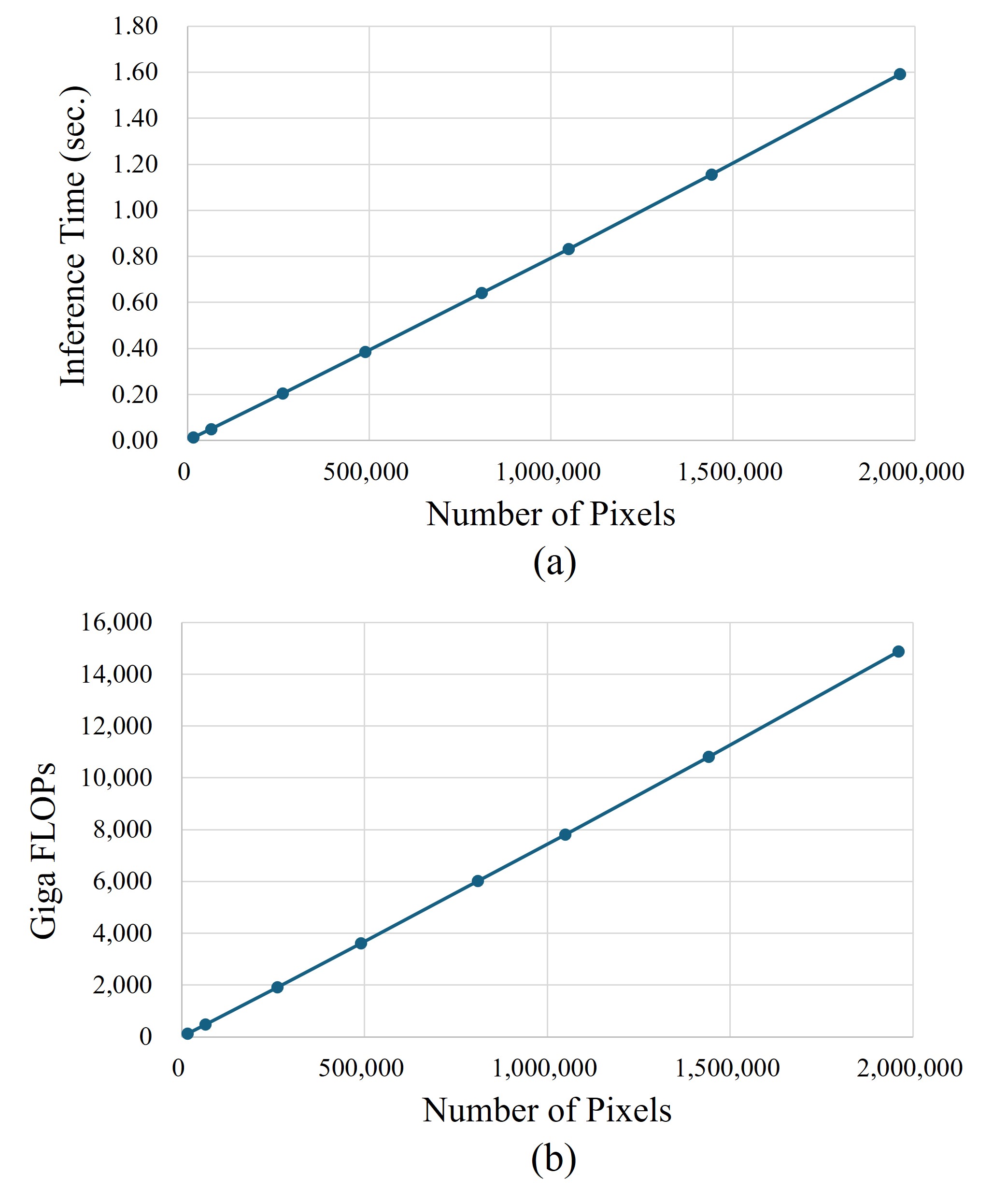}
\caption{{The scalability analysis of the proposed ExplainS2A algorithm, including (a) the inference time measured in sec. and (b) the computational complexity measured in FLOPs.
Both inference time and FLOPs exhibit linear scaling with respect to the number of pixels, demonstrating the high efficiency and practical applicability of ExplainS2A for large-scale real-world applications.
Furthermore, ExplainS2A model is very lightweight, containing only 1,519,508 network parameters.}
}\label{fig:linear}
\end{figure}

\subsection{Scalability and Complexity Analysis}\label{sec:TimeAnalysis}

In this section, we further investigate the computational complexity and algorithmic scalability of the proposed ExplainS2A, especially when processing large images.
As most spectral super-resolution algorithms were developed to process $256\times 256$ images, we test two large ROIs in order to facilitate remote sensing applications (cf. Figure \ref{fig:Figure6}).
The first AVIRIS ROI is located across the Georgia–Florida border (GA/FL), USA, acquired on May 7th, 2017.
The second AVIRIS ROI is located in California (CA), USA, acquired on August 2nd, 2019.
Both ROIs are with mixed landscapes, where the first one contains 1090 $\times$ 1798 pixels over city/farmland/mountain areas, and the second one contains 1080 $\times$ 1812 pixels over mountain/coastline regions.
The corresponding Sentinel-2 images were obtained using the same way as described in the beginning of Section \ref{sec: experiment}.
We test the representative methods, including MST++ \cite{MST++}, MSFN \cite{MSFN}, SSU-Net \cite{SSU-Net}, COS2A \cite{COS2A}, as well as the proposed ExplainS2A algorithm.
The spatial quality of the HSIs reconstructed by these methods are summarized in Figure \ref{fig:Figure6}.
MSFN performs well in the CA region, but has some color distortion in the GA/FL border [cf. Figure \ref{fig:Figure6}(c)].
MST++ shows some spatially blurry results in the middle region of the GA/FL data, and has slight color distortion on the ocean surface in the CA data [cf. Figure \ref{fig:Figure6}(b)].
COS2A also shows slight color distortion on the ocean surface, while performing very well in the GA/FL area [cf. Figure \ref{fig:Figure6}(e)].
The SSU-Net [cf. Figure \ref{fig:Figure6}(d)] and the ExplainS2A [cf. Figure \ref{fig:Figure6}(f)] yield outstanding spatial reconstructions for the GA/FL and CA regions. 
The superiority can be attributed to their specific design strengths.
Specifically, SSU-Net explicitly addresses the heterogeneous land cover in large-scale images, while ExplainS2A leverages the spectral-spatial duality theory embedded in its fusion network design [cf. Figure \ref{fig:whole_model}(b)].

Next, let us investigate the spectral reconstruction quality.
The Figure \ref{fig:Figure7} displays the SAM error maps of the methods under test.
Given a testing ROI and a spectral super-resolution method in Figure \ref{fig:Figure7}, the SAM value of the $i$th pixel is computed as the angle error between the $i$th reconstructed hyperspectral pixel in $\bY_H^\star$ and the $i$th hyperspectral pixel in the real AVIRIS image $\bY_H$.
Thus, the SAM error maps characterize the spectral fidelity of the reconstructed pixels in the HSI $\bY_H^\star$.
One can observe that for the GA/FL border ROI, MST++ and MSFN both show much larger angle errors than SSU-Net, COS2A, and ExplainS2A, especially in the mountainous area (cf. Figure \ref{fig:Figure7}).
As for the ROI in the CA region, MSFN shows higher errors in the ocean surface area, while MST++ exhibits higher angle errors in the mountainous area (cf. Figure \ref{fig:Figure7}).
For both ROIs, SSU-Net, COS2A, and ExplainS2A demonstrate their strong hyperspectral reconstruction capability, as indicated by the very low SAM angle errors.
Notably, for the middle and lower parts of the CA data, ExplainS2A still demonstrates lower angle errors than COS2A (cf. Figure \ref{fig:Figure7}), well echoing the truly superior performance of ExplainS2A in Table \ref{tab: quan}.

Remarkably, while ExplainS2A achieves such a strong spectral super-resolution result, its computational time is much faster than other peer methods.
For COS2A, due to its reliance on NMF-based optimization, it requires approximately 4 hours to process each ROI, which is impractical for scenarios requiring real-time computing.
%
In contrast, SSU-Net demonstrates good computational efficiency, requiring 1.9038 and 1.9107 seconds (sec.) to process the two images, respectively.
Remarkably, ExplainS2A further reduces the inference time to 1.6863 and 1.6865 sec. when processing the large GA/FL image and the CA image, respectively, while achieving even better SAM and RMSE results.
MST++ is also computationally efficient, as it takes 1.9070 and 2.3133 sec. for the two images, respectively.
MSFN takes about 1.2 minutes for both images.
Notably, the proposed ExplainS2A achieves the desired linear computational complexity \cite{SLIC2012} with respect to the input size, as evaluated on sub-areas of the GA/FL ROI.
As demonstrated in Figure \ref{fig:linear}(b), the complexity measured in the floating point operations (FLOPs) also exhibits a linear scalability.
%
%
Specifically, for an image with about one million pixels, ExplainS2A completes the inference in approximately 0.8 sec. and takes just about 8000 Giga FLOPs.
This high computational efficiency is attributed to the lightweight design of the architecture (cf. Figure \ref{fig:whole_model}), which contains only about 1.5 million network parameters, thereby ensuring superior scalability for large-scale image processing (cf. Figure \ref{fig:linear}).

\begin{table}[t!]
\footnotesize
\centering
\caption{Quantitative comparison of different strategies (i.e., mathematical, hybrid, and learnable) for solving \eqref{prob:minimize}.}
%
\label{tab: ablation_unfolding}
\setlength{\tabcolsep}{1mm}
\renewcommand{\arraystretch}{1.2}
\scalebox{0.92}{
\begin{tabular}{c|ccccc}
\hline
\hline
Strategy (Strat.) & PSNR $\!(\uparrow)$ & SAM $\!(\downarrow)$ & RMSE $\!(\downarrow)$ & SSIM $\!(\uparrow)$ & Time (sec.) \\
\hline
\multirow{3}{*}{}
Mathematical Strat. & 39.0667 & 1.4868 & 0.0067 & 0.9866 & 9.2445\\
Hybrid Strat. & 39.3467 & 1.4921 & 0.0067 & 0.9868 & 2.3083\\
Learnable Strat. & \textbf{39.4216} & \textbf{1.4499} & \textbf{0.0066} & \textbf{0.9876} & \textbf{0.0475} \\
\hline
\hline
\end{tabular}}
\end{table}

\begin{table}[t!]
\footnotesize
\centering
\caption{Ablation study for the proposed ExplainS2A algorithm,
%
where the marker ``$\checkmark$'' denotes that the corresponding attention mechanism is used.}
\label{tab:ablation}
\setlength{\tabcolsep}{2mm}
\renewcommand{\arraystretch}{1.2}
\scalebox{0.94}{
\begin{tabular}{cc|cccc}
\hline
\hline
\multicolumn{2}{c|}{Attention Mechanism} & \multicolumn{4}{c}{Metrics} \\
\cline{1-2} \cline{3-6} 
Spectral & Spatial & PSNR $\!(\uparrow)$ & SAM $\!(\downarrow)$ & RMSE $\!(\downarrow)$ & SSIM $\!(\uparrow)$ \\
\hline
      &       & 38.6412 & 1.7007 & 0.0073 & 0.9856 \\
\checkmark &       & 39.3479 & 1.4500 & \textbf{0.0066} & 0.9874 \\
      & \checkmark & 38.7018 & 1.7007 & 0.0073 & 0.9857 \\
\checkmark & \checkmark & \textbf{39.4216} & \textbf{1.4499} & \textbf{0.0066} & \textbf{0.9876} \\
\hline
\hline
\end{tabular}}
\end{table}

\subsection{Discussions}\label{sec:discussions}

In this section, we first investigate the necessity of employing a deep network to solve \eqref{prob:minimize}.
By modifying the learnable deep unfolding network within ExplainS2A [cf. Fig. \ref{fig:whole_model}(a)], we derive two additional variants for comparison, including the mathematical variant and the hybrid variant.
For the mathematical variant, all subproblems are solved directly through mathematical operators [i.e., the closed-form solutions for \eqref{prob:Z-Update} and \eqref{prob:Y_H-Update}, and the dual variable update], and the deep learning-based $\text{REG}^{\text{I}}$ is replaced by the $\text{REG}^{\text{E}}_{\text{spec}}$, which is implemented using the split Bregman method\footnote{ The split Bregman method: \url{https://www.mathworks.com/matlabcentral/fileexchange/36278-split-bregman-method-for-total-variation-denoising}.}.
Accordingly, the subsequent deep fusion stage is trained using the following loss function, i.e., $\|\bY_H^\star-\bY_H\|_1 + \lambda~\! \textrm{REG}^{\text{E}}_{\text{spat}}(\bY_H^\star)$. 
Furthermore, as $\bD$ cannot be adaptively estimated within the mathematical solver, we pretrain it through a single convolution layer.
Regarding the hybrid variant, only \eqref{prob:Z-Update} is solved using $\text{REG}^{\text{E}}_{\text{spec}}$ in a mathematical manner, while the remaining subproblems [i.e., \eqref{prob:Y_H-Update} and the dual variable update] are optimized in a learnable manner.
Accordingly, the loss function is designed as $\|\widetilde{\bY}_H-\bY_H\|_1 + \|\bY_H^\star-\bY_H\|_1 + \lambda \textrm{REG}^{\text{E}}_{\text{spat}}(\bY_H^\star)$.
As shown in Table \ref{tab: ablation_unfolding}, under the same $K=4$ stages, the mathematical and hybrid variants exhibit a degradation in quantitative performance, while ExplainS2A, with its high-flexibility learnable unfolding network, achieves superior performance with only a few stages.
In addition, in terms of computational time, ExplainS2A is about two orders of magnitude faster than the other two variants, demonstrating the computational advantage of the learnable strategy.

Furthermore, we analyze the contribution of the specific attention mechanisms (i.e., spectral attention and spatial attention) designed in Section \ref{sec:algodesign}.
Specifically, by replacing the attention weights with their global average, we can eliminate the influence of the adaptive attention mechanism without changing the underlying feature space.
As shown in Table \ref{tab:ablation}, when the spectral attention is used, the model achieves a low SAM score compared to the baseline, highlighting its specific contribution to preserving spectral fidelity. 
On the other hand, the spatial attention mechanism primarily ensures spatial structural consistency, thereby enhancing spatial reconstruction metrics (i.e., PSNR and SSIM).
By simultaneously using both attention mechanisms, the model achieves superior performance across all metrics.

\begin{figure}[t]
    \centerline{\includegraphics[width=0.5\textwidth]{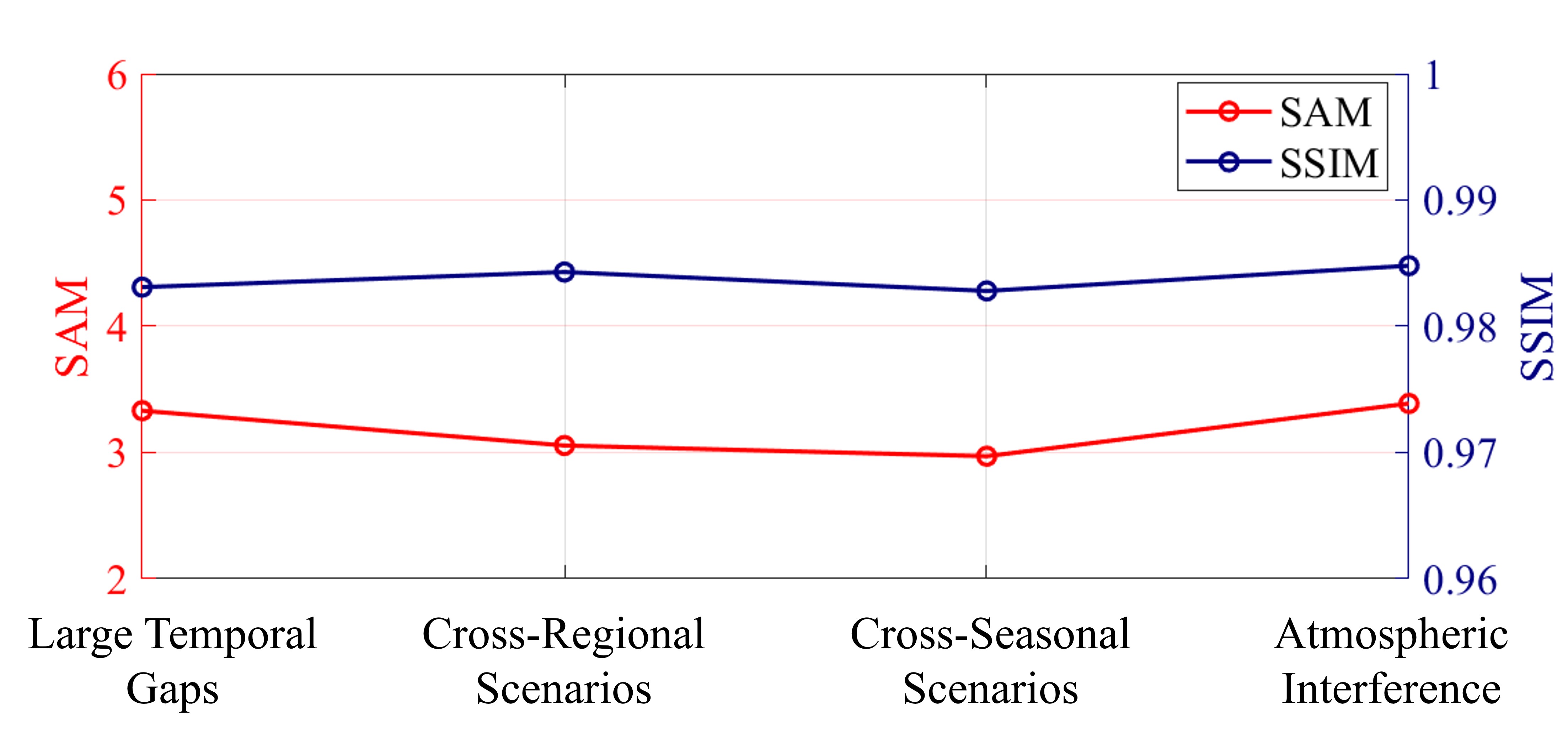}}
    \caption{Stable performances of ExplainS2A across diverse challenging conditions. 
    SAM is used to assess spectral structure similarity, while SSIM measures spatial structure similarity.
    }\label{fig: conditions}
\end{figure}

To comprehensively evaluate the robustness of ExplainS2A algorithm, we conduct the experiments under four distinct and challenging conditions: large temporal gaps, cross-regional scenarios, cross-seasonal scenarios, and atmospheric interference (aerosol and water vapor).
For the large temporal gap condition, we select test images acquired in 2016, 2017, and 2024, creating a significant time interval from the training data in 2018 to verify robustness against long-term land-cover changes.
Regarding the cross-regional condition, the training data are collected from the Western USA, Eastern USA, and Canada regions, and we further select the testing data from the westernmost (i.e., Hawaii, USA) and central USA areas to maximize spatial diversity. 
For the cross-seasonal condition, the model trained on summer data (June) is now evaluated on test data collected in winter (December and January), thereby introducing substantial seasonal variations.
Finally, for the atmospheric interference condition, we specifically select testing samples characterized by natural aerosols, such as cloud, fog, and mist, to assess performance under atmospheric scattering and absorption effects.
For each scenario, we select 7 to 14 samples to ensure a representative analysis, and the averaged quantitative results are summarized in Figure \ref{fig: conditions}.
As illustrated in Figure \ref{fig: conditions}, the proposed ExplainS2A demonstrates consistent and good performance across various challenging conditions.
Specifically, it maintains the SAM values to be less than 4 degrees, with the SSIM values consistently exceeding 0.98.
These findings highlight the robustness of the ExplainS2A algorithm, verifying its generalizability and potential for practical scenarios.


\begin{table}[t!]
\footnotesize
\centering
\caption{Systematic comparison between COS2A and ExplainS2A. 
The number of trainable parameters of COS2A involves only its deep learning part because the convex optimization part of COS2A does not require any training.}
\label{tab:comparison}
\setlength{\tabcolsep}{1.5mm}
\renewcommand{\arraystretch}{1.35}
\scalebox{0.94}{
\begin{tabular}{c|c|c}
\hline
\hline
\multirow{2}{*}{Features} & \multicolumn{2}{c}{Methods}\\
\cline{2-3}
& COS2A \cite{COS2A} & ExplainS2A\\
\hline
\multirow{2}{*}{Ingredient}& Deep Unfolding & Deep Unfolding \\
& Coupled-NMF & Deep Fusion
\\
\hline
\multirow{3}{*}{Regularizer} & DIP & DIP\\
& Spectrum Simplex Volume & Spectral TV \\
& Abundance Sparsity & Spatial TV \\
\hline
Trainable Parameters & 746,662 & 1,519,508\\
\hline
Training Time & $\approx$ 11.5 hours & $\approx$ 12.5 hours\\
\hline
Inference Time & $\approx$ 14.64 sec. & $\approx$ 0.05 sec.\\
\hline
Complexity & $\propto L $ \cite[Appendix B]{COS2A} & $\propto L $ (cf. Figure \ref{fig:linear})\\
\hline
\multirow{2}{*}{Required Prior} & Blurring Kernel & \multirow{2}{*}{None}\\
& SRT Matrix & \\
\hline
Data Scarcity Sensitivity& Very Low (cf. Table \ref{tab:data_sensitivity_combined})& Low (cf. Table \ref{tab:data_sensitivity_combined})\\
\hline
\hline
\end{tabular}}
\end{table}

Given that COS2A \cite{COS2A} is the only peer method that exactly addresses our target problem, we present a systematic comparison between it and the proposed ExplainS2A across various aspects, as summarized in Table \ref{tab:comparison}.
From a mathematical perspective, both methods aim to minimize the distance between the spectrally downsampled HSI $\bY_H^\star$ and the input MSI.
However, their architectural strategies differ significantly.
COS2A relies on the CODE theory, utilizing a deep unfolding network to first estimate a rough LR HSI, followed by a mathematical NMF solver to fuse the LR HSI with the HR bands inherent in the MSI.
In contrast, ExplainS2A is an end-to-end deep learning framework.
It employs the same deep unfolding architecture for the initial spectral super-resolution but replaces the mathematical part with an explainable deep fusion network, enabling near-real-time reconstruction.
Regarding the regularization term (i.e., REG), both methods share the same deep image prior (i.e., $\textrm{REG}^{\text{I}}$) implicitly inherent in the deep unfolding network.
Besides, COS2A leverages the low-rank property of HSIs, decomposing the image into an endmember matrix and an abundance matrix, and applying minimum simplex volume and abundance sparsity regularizers, respectively \cite{COCNMF}.
Conversely, ExplainS2A explicitly learns the priors through $ \textrm{REG}^{\text{E}}_{\text{spec}}$ and $\textrm{REG}^{\text{E}}_{\text{spat}}$, which are for promoting the spectral and spatial smoothness, respectively. 
As the mathematical NMF part in COS2A is replaced by a deep fusion network, ExplainS2A contains more trainable parameters than COS2A.
However, under exactly the same setting, the training time of ExplainS2A is nearly the same as COS2A, as detailed in Table \ref{tab:comparison}.
In terms of inference complexity, both methods exhibit linear complexity with respect to the number of pixels $L$.
Nevertheless, 
as shown in Table \ref{tab:comparison}, ExplainS2A achieves the inference time of only about $0.05$ sec. for a $256 \times 256$ image, which is approximately 300 times faster than COS2A (about $14.64$ sec.).
Moreover, COS2A relies on two prior inputs, namely the blurring kernel and the SRT matrix, whereas ExplainS2A does not require any explicit prior information.
That said, ExplainS2A is able to blindly complete the Sentinel-to-AVIRIS transform.
All related measurements are conducted on the same hardware specifications detailed in Section \ref{sec:experimental setting}.

Finally, to investigate the sensitivity to data scarcity, we trained both COS2A and ExplainS2A using only 50\% of the original training data (reduced from 568 to 284 samples) and reported the quantitative results in Table \ref{tab:data_sensitivity_combined}.
As expected, the performances of both methods degrade.
However, they exhibit extremely low sensitivity to the data scarcity (COS2A, in particular). 
This is attributed to the CODE theory \cite{CODE}, where the deep unfolding network only provides a rough solution, and the subsequent convex optimization is responsible for extracting key information (by $\bQ$-quadratic-norm) to enhance the solution \cite{COS2A}.
On the other hand, it is crucial to notice that even with only half of the training samples, ExplainS2A (trained on 284 samples) still significantly outperforms COS2A (trained on 568 samples) in all quantitative metrics, demonstrating its superior capability in achieving the complicated spectral-spatial duality transform.

\begin{table}[t!]
\footnotesize
\centering
\caption{Data scarcity sensitivity analysis for COS2A \cite{COS2A} and ExplainS2A across different training dataset sizes.}
\label{tab:data_sensitivity_combined}
\setlength{\tabcolsep}{1.2mm}
\renewcommand{\arraystretch}{1.2}
\scalebox{0.94}{
\begin{tabular}{c|c|cccc}
\hline
\hline
\multirow{2}{*}{Dataset Size} & \multirow{2}{*}{Methods} & \multicolumn{4}{c}{Metrics} \\
\cline{3-6}
&  & PSNR $\!(\uparrow)$ & SAM $\!(\downarrow)$ & RMSE $\!(\downarrow)$ & SSIM $\!(\uparrow)$ \\
\hline
\multirow{2}{*}{284} & COS2A 
& 34.9022 & 2.6532 & 0.0120 & 0.9470\\
& ExplainS2A & \textbf{37.8709} & \textbf{1.8078} & \textbf{0.0077} & \textbf{0.9828} \\
\hline
\multirow{2}{*}{568} & COS2A & 35.0084 & 2.5037 & 0.0115 & 0.9488 \\
& ExplainS2A & \textbf{39.4216} & \textbf{1.4499} & \textbf{0.0066} & \textbf{0.9876}  \\
\hline
\hline
\end{tabular}}
\end{table}

\subsection{Case Study on Blind Source Separation}\label{sec:CaseStudy}

\begin{figure}[t]
\centerline{\includegraphics[width=0.5\textwidth]{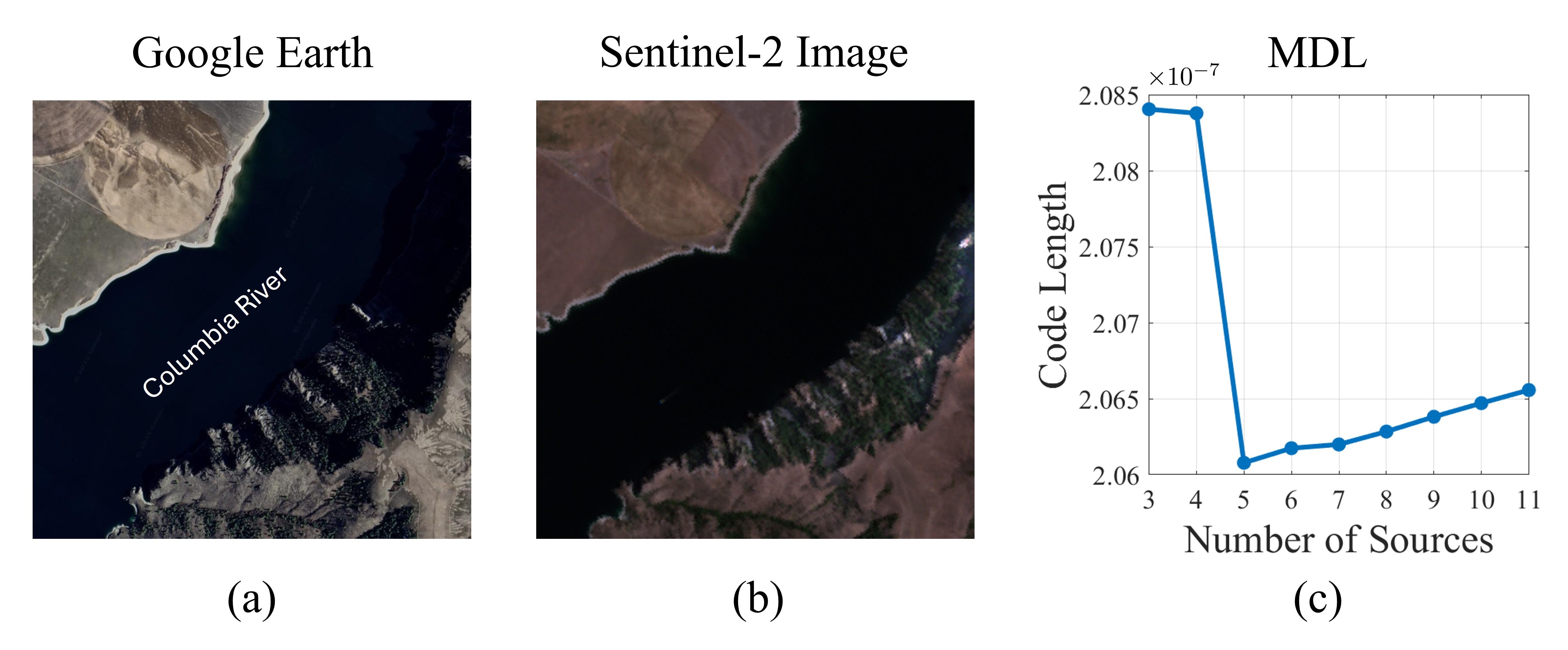}}
\caption{(a) ROI displayed by Google Earth imagery, (b) ROI displayed by Sentinel-2 satellite, and (c) the model-order selection for the studied ROI ($256\times 256$ pixels) by the minimum description length (MDL) criterion \cite{MDL}.
}\label{fig:Figure9}
\end{figure}

\begin{figure}[t]
\centerline{\includegraphics[width=0.5\textwidth]{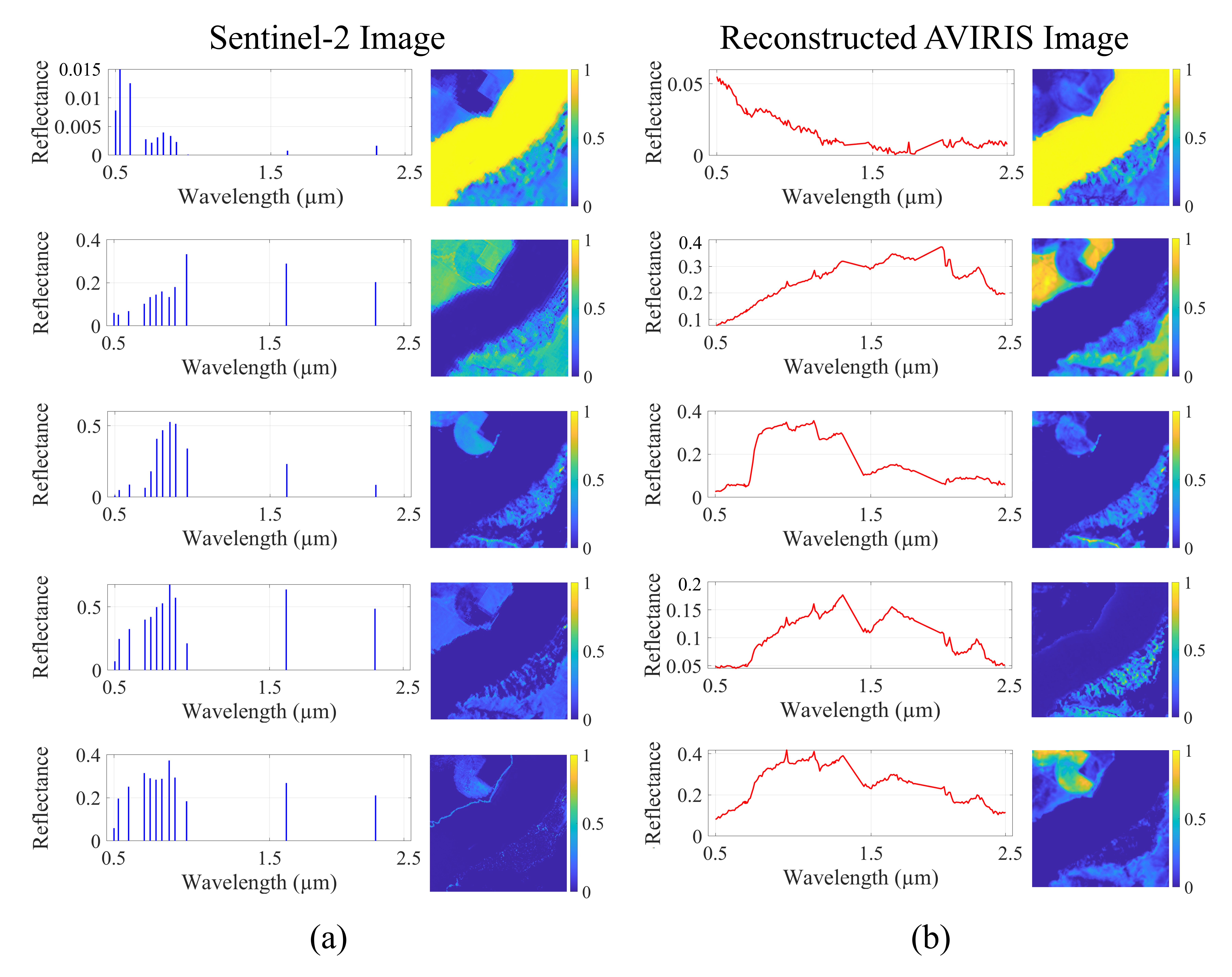}}
\caption{(a) BSS results obtained from the Sentinel-2 data $\bY_S$, including five multispectral endmembers (i.e., the blue pulses) and their corresponding abundances.
(b) BSS results obtained from the ExplainS2A-reconstructed AVIRIS data $\bY_H^\star$, including five hyperspectral endmembers (i.e., the red curves) and their corresponding abundances.
The model-order of $N:=5$ is determined by the MDL criterion [cf. Figure \ref{fig:Figure9}(c)].
}\label{fig:Figure10}
\end{figure}

Blind source separation (BSS) is a critical signal processing technology for understanding the identities of materials presented in a remotely sensed image and their spatial distribution maps (abundance maps) \cite{HyperCSI}.
BSS often relies on spectral information (e.g., those retained in HSI) in order to compute the endmember signatures for material identification \cite{VCA}.
Thus, performing BSS on MSI (e.g., Sentinel-2 image $\bY_S$) often leads to unsatisfactory results, such as repeatedly identified sources (cf. Figure \ref{fig:Figure10}(a)).
Hence, ExplainS2A is able to turn the Sentinel-2 MSI $\bY_S$ into its counterpart HSI $\bY_H^\star$, thereby facilitating the effectiveness of BSS, as will be demonstrated in this section.

To demonstrate the strong identifiability of the ExplainS2A-reconstructed AVIRIS image, which should be attributed to the high spectral fidelity of $\bY_H^\star$ (cf. Table \ref{tab: quan} and Figure \ref{fig:Figure7}), we conduct a BSS case study based on the ROI located in the Washington (WA), USA \cite[Table 1]{COS2A}.
This ROI is with mixed landscapes, containing $256 \times 256$ pixels over river/farmland/mountainous areas. 
The ROI is illustrated by Google Earth data \cite{GoogleEarth} [cf. Figure \ref{fig:Figure9}(a)] and Sentinel-2 data in August 2019 [cf. Figure \ref{fig:Figure9}(b)] for comprehensively analyzing the unmixed sources.
To estimate the number of sources (substances/materials) presented in the ROI, we adopt a fully-unsupervised and parameter-tuning-free algorithm, called minimum description length (MDL) \cite{MDL}.
The MDL algorithm is developed based on information theory, and estimates the number of sources as the model-order $N$ that yields the shortest code length when coding/describing the real HSI $\bY_H$; mathematical details of deriving the code length can be found in \cite{MDL}.
Accordingly, the number of sources is estimated to be $N:=5$ based on MDL, as detailed in Figure \ref{fig:Figure9}(c).

Subsequently, we use the vertex component analysis (VCA) algorithm \cite{VCA} to extract $N:=5$ multispectral endmembers from the Sentinel-2 image $\bY_S$, followed by using the fully constrained least-squares (FCLS) algorithm \cite{FCLS} to derive their associated abundance maps, as presented in Figure \ref{fig:Figure10}(a).
Then, we also use the proposed ExplainS2A, trained using the same dataset provided in \cite[Table 1]{COS2A}, to convert $\bY_S$ into AVIRIS-level data $\bY_H^\star$.
Similarly, VCA is applied on $\bY_H^\star$ to extract $N:=5$ hyperspectral endmembers, followed by using FCLS to derive the associated abundance maps, as presented in Figure \ref{fig:Figure10}(b).

The first abundance map obtained from HSI $\bY_H^\star$ [cf. Figure \ref{fig:Figure10}(b)] accurately delineates the river channel, clearly highlighting its spatial extent.
In contrast, the first abundance derived from MSI $\bY_S$ [cf. Figure \ref{fig:Figure10}(a)] also identifies the main river region but exhibits an overestimated abundance toward the lower-right non-river region. 
%
%
In the second row of Figure \ref{fig:Figure10}(b), the second abundance map from HSI $\bY_H^\star$ successfully isolates the sandy regions [cf. Figure \ref{fig:Figure9}(a)], achieving a distinct separation between different land-cover types.
Conversely, the second Sentinel-2 abundance map [cf. Figure \ref{fig:Figure10}(a)] still exhibits a noticeable mixing effect, where portions of the circular vegetated area in the upper-left corner [cf. Figure \ref{fig:Figure9}(a)] are incorrectly interpreted as sandy regions.
As for the third source, the abundance maps unmixed from both $\bY_H^\star$ and $\bY_S$ effectively distinguish the vegetated regions located in the upper-left and lower-right areas of the scene (cf. Figure \ref{fig:Figure9}).
However, the fourth Sentinel-2 abundance map exhibits a distribution pattern highly similar to the second Sentinel-2 abundance map [cf. Figure \ref{fig:Figure10}(a)].
%
The phenomenon of repeatedly identified sources (though with weaker abundances) mainly stems from the insufficient number of spectral bands in typical MSI, easily resulting in indistinguishable endmembers that further cause the ill-conditioned BSS scenario \cite{HISUN}.
%
%
In contrast, the fourth AVIRIS abundance map [cf. Figure \ref{fig:Figure10}(b)] clearly distinguishes the rocky materials located on the cliff area in the lower-right region (cf. Figure \ref{fig:Figure9}).
This well substantiates that the spectrally super-resolved HSI $\bY_H^\star$ provides richer and more discriminative spectral signatures, thereby enhancing source separability.
Finally, the fifth AVIRIS abundance map [cf. Figure \ref{fig:Figure10}(b)] distinctly separates the circular farmland region located in the upper-left area (cf. Figure \ref{fig:Figure9}), demonstrating a clear boundary between crop and non-crop areas.
Although the fifth Sentinel-2 abundance map also captures this circular farmland region, its material distinction is notably less pronounced, as indicated by the faint abundance map [cf. Figure \ref{fig:Figure10}(a)].
The easily distinguishable hyperspectral endmembers [i.e., the red curves in Figure \ref{fig:Figure10}(b)] contribute to the strong BSS results achieved by the ExplainS2A-reconstructed HSI $\bY_H^\star$.

\section{Conclusions and Future Works}\label{sec:conclusion}

In this work, we investigated the highly ill-posed inverse problem of spectrally super-resolving the 12-band multi-resolution Sentinel-2 image to a 172-band AVIRIS-level hyperspectral image of uniformly high 10-m spatial resolution.
This technique allows those historical Sentinel-2 data to be computationally converted into their corresponding hyperspectral counterparts, directly contributing to advanced remote sensing data analysis and applications.
For example, blind source separation (BSS) is a critical data analysis tool for computing the abundance distribution maps of the underlying substances/materials presented in an ROI, while the band insufficiency of typical MSI satellites (e.g., Sentinel-2) make BSS less effective as the multispectral signatures would be too similar for distinct substances/materials (cf. Section \ref{sec:CaseStudy}).
This motivates us to transform the widely available Sentinel-2 data into hyperspectral data to facilitate BSS and the subsequent identification/classification missions.

Although the 12-to-172 conversion is too difficult even for those seminal spectral super-resolution methods \cite{MST++}, this ambitious yet critical aim had been achieved once before by COS2A \cite{COS2A}.
We go a step further to propose the ExplainS2A algorithm based on the spectral-spatial duality theory, resulting in an explainable, model-order estimation-free, very lightweight, and much faster algorithm (more than 300 times faster than COS2A; cf. Table \ref{tab: quan}).
%
ExplainS2A is developed based on deep unfolding theory and explainable spatiospectral feature fusion, and contains only about 1.5 million network parameters (cf. Figure \ref{fig:whole_model}).
%
The hyperspectral images reconstructed by ExplainS2A have superior spatial quality (as indicated by the outstanding PSNR/SSIM values; cf. Table \ref{tab: quan}) and high spectral fidelity to real AVIRIS data (as indicated by the SAM error maps; cf. Figure \ref{fig:Figure7}) over diverse landscapes (cf. Section \ref{sec:TimeAnalysis}).
ExplainS2A achieves state-of-the-art spectral super-resolution performance for the aforementioned critical remote sensing problem with linear computational complexity, and is able to fast super-resolve large-scale Sentinel-2 data (e.g., 0.8 sec. for million-scale image) while maintaining high-precision hyperspectral reconstruction.

In future research, the ExplainS2A methodology can be leveraged by the remote sensing community to enable hyperspectral missions—particularly those necessitating high substance/material discriminability for distinguishing remote targets—within areas monitored by the Sentinel-2 satellite.
Furthermore, the exploration of quantum-based real-time implementations (e.g., quantum generative model \cite{HyperKING} or quantum light-splitting prism \cite{PRIME}), as well as the development of energy-efficient hardware architectures for ExplainS2A, represent promising directions for enabling its deployment in next-generation onboard edge-computing platforms.

\bibliography{ref}

\begin{IEEEbiography}[{\resizebox{0.9in}{!}{\includegraphics[width=1in,height=1.25in,clip,keepaspectratio]{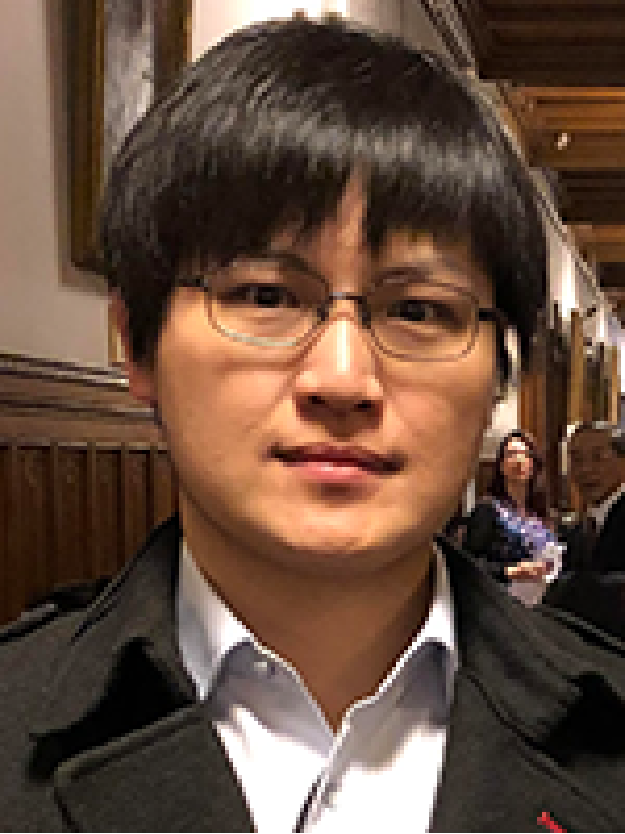}}}]
{\bf Chia-Hsiang Lin}
(S'10-M'18-SM'24)
received the B.S. degree in electrical engineering and the Ph.D. degree in communications engineering from National Tsing Hua University (NTHU), Taiwan, in 2010 and 2016, respectively.
From 2015 to 2016, he was a Visiting Student of Virginia Tech,
Arlington, VA, USA.

He is currently a Professor with the Department of Electrical Engineering,
National Cheng Kung University (NCKU), Taiwan, and also serves as a Technical Director of Smart Sensing \& Systems Technology Center, Industrial Technology Research Institute (ITRI).
Before joining NCKU, he held research positions with The Chinese University of Hong Kong, HK (2014 and 2017),
NTHU (2016-2017),
and the University of Lisbon (ULisboa), Lisbon, Portugal (2017-2018).
He was an Assistant Professor with the Center for Space and Remote Sensing Research, National Central University, Taiwan, in 2018, a Visiting Professor with ULisboa, in 2019, and a Visiting Professor with Texas A\&M University, USA, in 2025.
His research interests include network science,
quantum computing,
convex geometry and optimization, blind signal processing, and imaging science.

Dr. Lin received the Emerging Young Scholar Award (The 2030 Cross-Generation Program) from National Science and Technology Council (NSTC), from 2023 to 2027,
the Future Technology Award from NSTC, in 2022,
the Outstanding Youth Electrical Engineer Award from The Chinese Institute of Electrical Engineering (CIEE), in 2022,
the Best Young Professional Member Award from IEEE Tainan Section, in 2021,
and the Prize Paper Award from IEEE Geoscience and Remote Sensing Society (GRS-S), in 2020.
He received the Ministry of Science and Technology (MOST) Young Scholar Fellowship, together with the EINSTEIN Grant Award, from 2018 to 2023.
In 2016, he was a recipient of the Outstanding Doctoral Dissertation Award from the Chinese Image Processing and Pattern Recognition Society and the Best Doctoral Dissertation Award from the IEEE GRS-S.
\end{IEEEbiography}

\begin{IEEEbiography}[{\resizebox{1in}{!}{\includegraphics[width=1in,height=1.25in,clip,keepaspectratio]{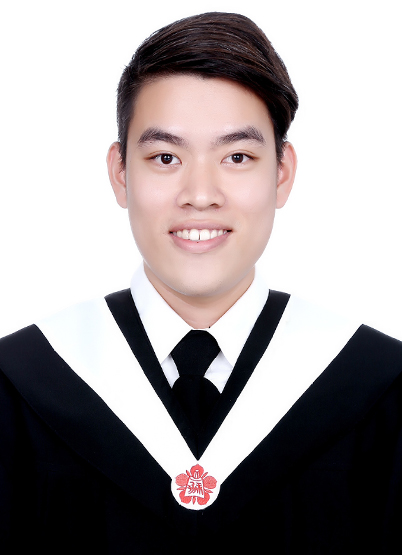}}}]
{\bf Zi-Chao Leng}
received his B.S. degree from the Department of Electronic Engineering, National Cheng Kung University, Taiwan, in 2021.

He is currently a Ph.D. student with Intelligent Hyperspectral Computing Laboratory, Institute of Computer and Communication Engineering, National Cheng Kung University, Taiwan. 
His research interests include deep learning, convex optimization, hyperspectral imaging, and biomedical imaging.
\end{IEEEbiography}

\end{document}